\documentclass[aps,pra,twocolumn,10pt,showpacs,superscriptaddress,amsmath,amssymb,amsfonts,floatfix,longbibliography]{revtex4-2}

\usepackage[T1]{fontenc} 
\usepackage{graphicx}
\usepackage{float}
\usepackage{braket}
\usepackage{dcolumn}
\usepackage{bm}
\usepackage[normalem]{ulem}
\usepackage{amssymb}
\usepackage{microtype}
\usepackage{xfrac}
\usepackage{gensymb}
\usepackage[most]{tcolorbox}
\usepackage{xcolor}
\usepackage{multirow}
\usepackage{enumitem}
\usepackage{natbib,hyperref}
\usepackage{graphicx}
\usepackage{booktabs}
\usepackage{chemformula}
\usepackage{hyperref} 

\hypersetup{
	colorlinks=true, 
	citecolor=blue,  
}

\graphicspath{{Figures/}{\main/Figures/}}

\makeatletter
\def\maketitle{
  \@author@finish
  \title@column\titleblock@produce
  \suppressfloats[t]
}
\makeatother



\begin{document}

\title{Rare-earth oxysulfides \ch{RE2O2S} as model mixed-anion frustrated magnets}

\author{Austin M. Ferrenti}
\thanks{austin.ferrenti@ubc.ca, alannah.hallas@ubc.ca}
\affiliation{Stewart Blusson Quantum Matter Institute, University of British Columbia, Vancouver, BC, V6T 1Z4, Canada}
\affiliation{Department of Physics \& Astronomy, University of British Columbia, Vancouver, BC, V6T 1Z1, Canada}

\author{Ksenia Khoroshun}
\affiliation{Stewart Blusson Quantum Matter Institute, University of British Columbia, Vancouver, BC, V6T 1Z4, Canada}
\affiliation{Department of Physics \& Astronomy, University of British Columbia, Vancouver, BC, V6T 1Z1, Canada}

\author{Mohamed Oudah}
\affiliation{Stewart Blusson Quantum Matter Institute, University of British Columbia, Vancouver, BC, V6T 1Z4, Canada}
\affiliation{Department of Physics \& Astronomy, University of British Columbia, Vancouver, BC, V6T 1Z1, Canada}

\author{Graham King}
\affiliation{Canadian Light Source, Saskatoon, SK, S7N 2V3, Canada}

\author{Jonathan Gaudet}
\affiliation{NIST Center for Neutron Research, National Institute of Standards and Technology, Gaithersburg, MD, 20899, United States}
\affiliation{Department of Materials Science and Eng., University of Maryland, College Park, MD, 20742-2115, United States}

\author{Alannah M. Hallas}
\thanks{austin.ferrenti@ubc.ca, alannah.hallas@ubc.ca}
\affiliation{Stewart Blusson Quantum Matter Institute, University of British Columbia, Vancouver, BC, V6T 1Z4, Canada}
\affiliation{Department of Physics \& Astronomy, University of British Columbia, Vancouver, BC, V6T 1Z1, Canada}
\affiliation{Canadian Institute for Advanced Research (CIFAR), Toronto, ON, M5G 1M1, Canada}

\begin{abstract}
\textbf{Abstract:}~The study of low-dimensional and anisotropic magnetism is a promising avenue for the discovery of novel quantum phenomena. Mixed-anion materials, defined by the coordination of metal cations by two or more distinct anionic species, provide an intrinsically anisotropic platform for the tuning of structural, magnetic, and electronic properties. In this work, we report the synthesis and characterization of a family of rare-earth oxysulfide (\ch{RE2O2S}) antiferromagnets (AFM) possessing a triangular-bilayer slab lattice geometry. Transitions consistent with long-range AFM order are observed for the majority of compositions, with several (RE = Ce, Pr, Nd, Sm) having been previously unreported. Although assumed to form stoichiometrically, pair distribution function analysis provides evidence for significant, synthesis-dependent interslab structural disorder as \ch{RE2O_{2+x}S_{1-x}}, resulting in significant variability in the bulk magnetic response of the Nd member. This work highlights the tunability of frustrated magnetic ground states in rare-earth-based mixed-anion materials, and the importance of thorough structural characterization in the discovery of new mixed-anion magnets.
\end{abstract}

\maketitle

\section*{Introduction}
Structurally anisotropic materials often possess enhanced magnetic and electrical properties, owing to electronic confinement effects \cite{rudenko2024anisotropic}. This anisotropy can be directly engineered via the substitution of entire anionic layers within a crystal structure, creating so-called mixed-anion materials defined by the anisotropic coordination of metal cations with two or more anionic species \cite{kageyama2018expanding}. Such materials have been largely underexplored, relative to single-anion systems, owing to the increased difficulty in obtaining structurally-ordered samples in a more chemically complex phase space \cite{miyoshi2021recent,katsumata2023development}. However, the introduction of this asymmetry affords incredible tunability of overall structural dimensionality, crystal electric field (CEF) energetics, and band gap width while maintaining overall structural rigidity \cite{kageyama2018expanding}.

In the realm of frustrated magnetism, mixed anions provide a route to the controlled design of novel structures with particular connectivity and magnetic dimensionality. For instance, the mineral-inspired quantum spin liquid candidate herbertsmithite, ZnCu$_3$(OH)$_6$Cl$_2$, achieves highly two-dimensional exchange interactions, which are mediated in-plane by hydroxide ligands, with negligible out-of-plane exchange through the interlayer chlorine~\cite{shores2005structurally}. Another example comes from the van der Waals antiferromagnet CrSBr, where the very different chemical characters of the sulfide and bromide anions produce highly anisotropic exchange interactions, yielding a complex magnetic ground state~\cite{lee2021magnetic,lopez2022dynamic}. Mixtures of anions can also be employed to modify cation valence states and thereby alter their magnetic character, as achieved in the topotactic nitridation of a molybdenum oxide pyrochlore to modify its spin from $S~=~1$ to $S=\sfrac{1}{2}$~\cite{clark2014spin}.

Particularly in the case of rare-earth-based magnets, the introduction of mixed-anion character enables fine control over the often complex CEF excitation spectrum, which dictates single-ion anisotropy, pseudo-spin state, and magnetic ground state stability. Until recently, literature reports of rare-earth mixed-anion materials have primarily focused on the impact of anisotropic coordination on their structural diversity and optical and luminescent properties \cite{orr2022structural}. However, several studies have highlighted the 
potential for interesting magnetism in certain rare-earth oxysulfides, \ch{RE2O2S} (RE~=~La~-~Lu), whose structure consists of staggered triangular-bilayer slabs of RE$^{3+}$ cations bridged by O$^{2-}$ and S$^{2-}$ anions within and between each bilayer, respectively \cite{quezel1970proprietes,suponitskii1988lanthanide,biondo2014geometric}. This structure represents a quasi-two-dimensional intermediate between the three-dimensional \ch{La2O3}-type and the fully two-dimensional triangular-lattice antiferromagnets (AFMs), as realized in delafossite-type materials, both of which have seen extensive study in recent years due to their ability to stabilize intrinsic magnetic frustration \cite{rai2020magnetism,xing2019crystal}.

\begin{figure*}[ht]
    \includegraphics[width=1\textwidth]{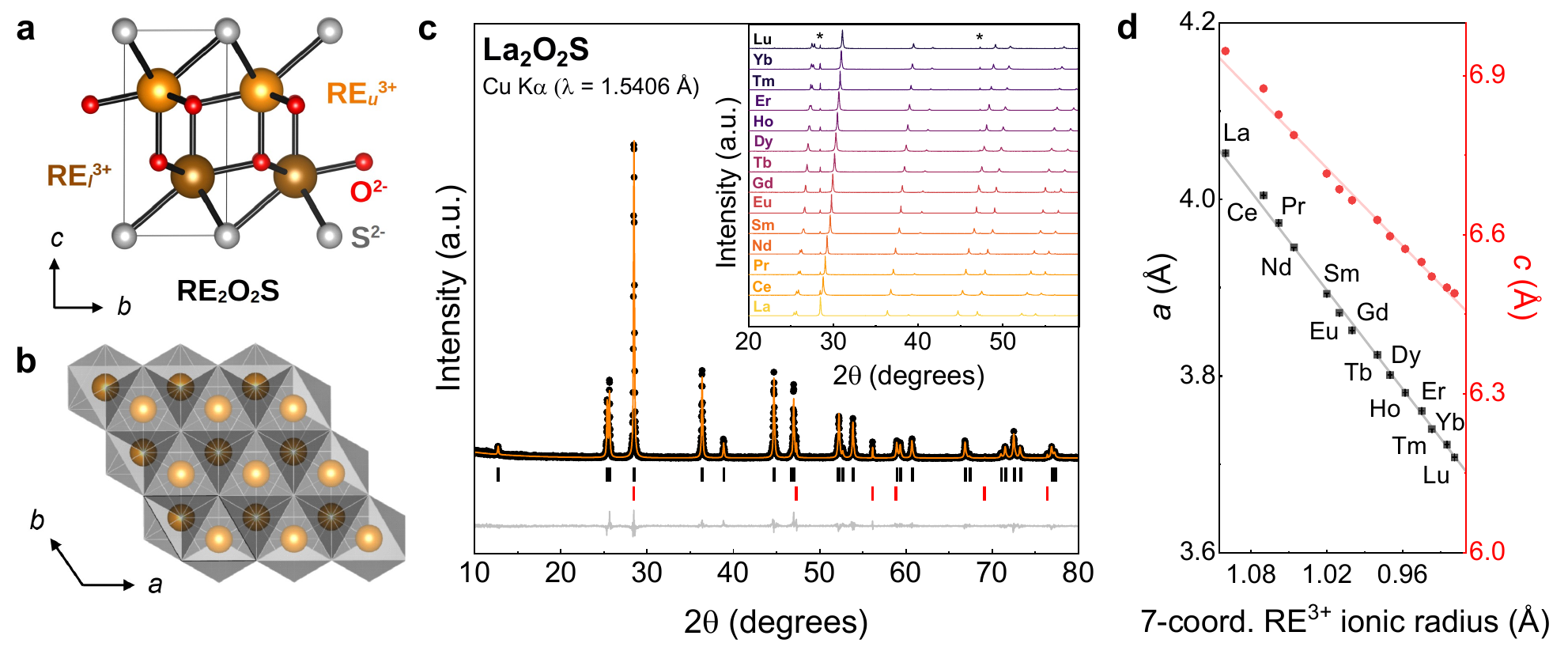}
    \caption{Crystal structure of the \ch{RE2O2S} family of mixed-anion magnets, shown (a) within and (b) normal to the triangular-bilayer slab. Although equivalent by symmetry, RE atoms in the upper (RE$_u^{3+}$) and lower (RE$_l^{3+}$) halves of the bilayer slab have been shaded differently for clarity. (c) Representative Rietveld refinement of powder X-ray diffraction (pXRD) data for \ch{La2O2S}, where the black, orange, and gray datasets denote the experimental and modeled patterns, and the difference between them. The black and red tick marks represent reflections attributed to \ch{La2O2S} and an internal \ch{Si} standard, respectively. The pXRD patterns for the full lanthanide series are shown in the inset with asterisks marking the Si standard. (d) Refined $a$ and $c$ lattice parameters for the trigonal \ch{RE2O2S} unit cell as a function of 7-coordinate RE$^{3+}$ ionic radius.}
    \label{Struc_comb}
\end{figure*}

In this work, we investigate the effect of mixed-anion anisotropy on magnetic ground states across the \ch{RE2O2S} family of AFMs possessing the \ch{La2O3}-type structure. In contrast to previous reports, the triangular-bilayer lattice geometry is found to stabilize long-range magnetic order at low temperatures for all but RE~=~Eu, Er, and Tm, suggesting a greater degree of structural order in samples grown by the method described in this work. Powder magnetization and specific heat measurements are in good agreement with the calculated CEF spectrum for each composition, revealing a diversity of single-ion anisotropies. The resulting magnetic phase diagrams are strongly influenced by these anisotropies, with many members of the \ch{RE2O2S} family exhibiting metamagnetic transitions. We also report strong synthesis-dependence in the bulk magnetic response. Pair distribution function (PDF) analysis of representative \ch{Nd2O2S} compositions grown by three different methods suggests that this variability is due to disorder largely localized on the sulfur site, likely through partial oxygen substitution. This work highlights the utility of mixed-anion substitution in tuning single-ion properties, as well as the synthetic limitations of such materials.
 
\section{Results and Discussion}

\subsection{Structure}

Many \ch{RE2O2S} synthesis techniques in the literature proceed through the direct reaction of either the rare-earth oxide (\ch{RE2O3}) and sulfur (S) or the rare-earth oxide and corresponding sulfide (\ch{RE2S3}) in an equimolar ratio~\cite{larquet2020metal,suponitskii1988lanthanide}. In this work, we employed the reaction of rare-earth metal, rare-earth oxide, and sulfur in a stoichiometric ratio, which was found to result in shorter reaction times and smaller oxide impurities in as-synthesized powder samples. While the A-type \ch{La2O3} structure ($P\overline{3}m1$ space group, no. 164) is only stable for the largest rare earths (RE~=~La~-~Pm) at ambient pressure~\cite{zinkevich2007thermodynamics}, the substitution of \sfrac{1}{3} of coordinating O$^{2-}$ anions for S$^{2-}$ extends the trigonal phase stability to the full lanthanide block. The \ch{RE2O2S} structure is comprised of hemispherically-coordinated RE$^{3+}$ cations arranged in an overall staggered triangular-bilayer slab geometry, a quasi-two-dimensional intermediate between the three-dimensional \ch{La2O3}-type structure and the conventional triangular lattice (Figure \ref{Struc_comb}a,b). Each cation is four-fold coordinated to O$^{2-}$ anions within the slab and three-fold coordinated to S$^{2-}$ anions between slabs, producing a highly anisotropic coordination environment. 

Powder X-ray diffraction (pXRD) patterns collected from \ch{RE2O2S} powder samples display a progressive rightward shift upon decreasing RE size, indicative of a shrinking of the \emph{P$\overline{3}$m1} (no. 164) unit cell, in good agreement with previous literature reports (Figure \ref{Struc_comb}c) \cite{ballestracci1968etude}. Rietveld refinement of the measured pXRD patterns yields an 8.5(1)\% and 6.6(1)\% decrease in the \textit{a} and \textit{c} lattice parameters from RE~=~La~-~Lu, respectively (Figure~\ref{Struc_comb}d), as well as a gradual shortening of all unique RE~-~RE, RE~-~O, and RE~-~S bonds (Tables S2, S3). All compositions additionally exhibit a strong, non-uniform preferred orientation which differs from that reported previously for polycrystalline \ch{RE2O2S} and could not be reduced via prolonged grinding, dilution with amorphous \ch{SiO2}, sifting, or increasing the sample volume \cite{zachariasen1949crystal,liu2018upconversion,sang2020synthesis}. These effects are most pronounced (but not limited to) reflections with a large \textit{l} component, owing to RE$_z$ and O$_z$ being the only refineable positions in the average structure. Air and overall structural stability are observed to decrease across the lanthanide series, with the measured pXRD of the smallest members of the series (RE~=~Tm~-~Lu) indicating the presence of \ch{RE2O3} and/or \ch{RE2S3} impurity phases after air exposure beyond 5-10 minutes. 

RE-RE distances in \ch{RE2O2S} phases are found to be intermediate between those observed in three-dimensional \ch{RE2O3} and purely two-dimensional \ch{AREO2} delafossite materials, in agreement with the previous literature \cite{eick1958preparation}. Taking \ch{Nd2O2S} as an example, the Nd-Nd distances within each triangular layer (\emph{i.e.} RE$_u$-RE$_u$ and RE$_l$-RE$_l$ in Figure~\ref{Struc_comb}a) are 3.9457(1)~\AA, comparable to the 3.831~\AA~and 3.622~\AA~in \ch{Nd2O3} and \ch{KNdO2}, respectively. The distance across the bilayer slab in \ch{NdO2S} (\emph{i.e.} RE$_u$-RE$_l$) is 3.765(5)~\AA, compared to 3.737~\AA~in \ch{Nd2O3}. Finally, the distance between the slabs is 4.422(5)~\AA, placing the \ch{RE2O2S} structure as a dimensional intermediate between \ch{KNdO2}, with interlayer spacings of 7.506~\AA, and \ch{Nd2O3}, with an inter-bilayer spacing 3.717~\AA, which is actually the shortest RE-RE distance in that structure. The substitution of the interlayer O$^{2-}$ in \ch{Nd2O3} with S$^{2-}$ results in only slight shifts in Nd-O-Nd bond angles within each triangular layer (106.0° to 104.3(7)°) and across the slab (112.7° to 114.1(6)°), as well as in the non-linear Nd-O/S-Nd bond angles between the slabs (91.7° to 83.5(1)° and 88.3° to 96.5(1)°). The average structural changes observed between the Nd member and \ch{Nd2O2S} are assumed to be representative across the rare-earth series.
 
Raman scattering spectra collected at room temperature from polycrystalline \ch{RE2O2S} samples also highlight the evolution of the \ch{La2O3}-type structure on decreasing RE size (Figure S2). With the exception of the Er and Yb members, all samples display only the four Raman-active optical modes (2$A_\text{1g}$ + 2$E_\text{g}$) allowed by the D$_{3D}$ point group symmetry of the \emph{P$\overline{3}$m1} (no. 164) space group. These correspond to RE-O vibrations within the triangular sublattices that comprise the overall bilayer slab, with the $E_\text{g}$ and $A_\text{1g}$ modes representing vibrations within the \textit{ab}-plane and along the \textit{c}-axis, respectively. All RE-S vibrations are expected to be active only in the far-infrared, limiting spectroscopic determination of potential interslab disorder. The lower-frequency modes, here denoted as $E_\text{g}^{\text{1}}$ and $A_\text{1g}^{\text{1}}$, are observed to gradually harden from RE = La - Lu. For RE = Ce, Pr, and Tb, these modes exhibit hardening beyond the overall trend. The higher-frequency modes, here denoted as $E_\text{g}^{\text{2}}$ and $A_\text{1g}^{\text{2}}$, instead soften upon decreasing RE size, with only RE = Ce and RE = Eu respectively displaying more significant softening and hardening. All deviations from the trends in vibrational mode frequencies notably occur for rare-earth cations possessing other stable valence states. Anharmonic behavior in materials such as these has been commonly attributed to abnormally strong interactions between phonons and crystal electric field levels. However, this may also arise as a result of higher levels of valence instability-driven disorder \cite{mcbride1994raman,dilawar2016pressure}.

\subsection{Magnetic properties}

\begin{figure*}[ht]
    \includegraphics[width=1.0\textwidth]{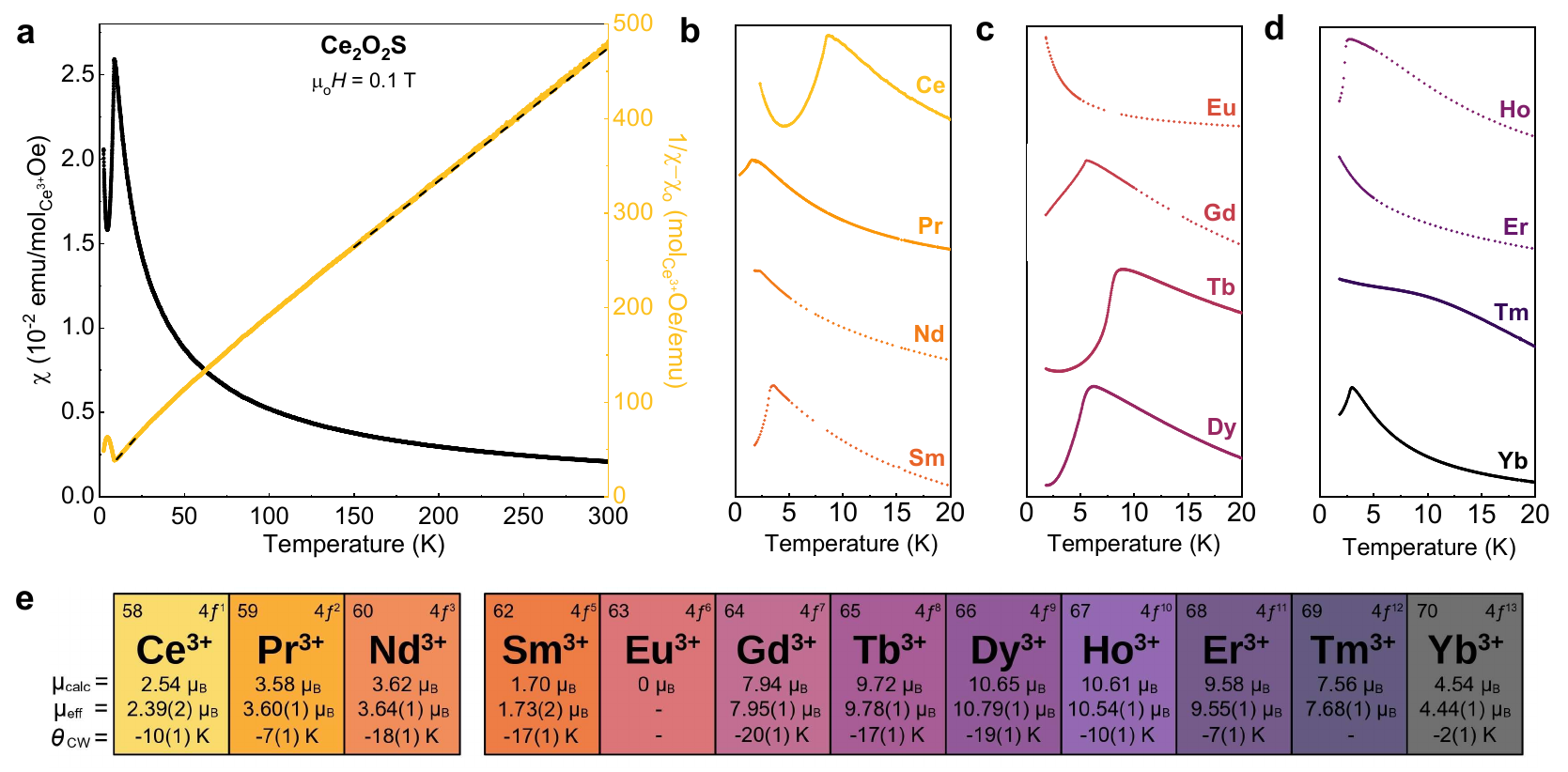}
    \caption{(a) Representative temperature-dependent magnetic susceptibility (black), its inverse (gold), and Curie Weiss fit lines (dashed black) of a \ch{Ce2O2S} powder sample. Scaled low-temperature magnetic susceptibility of representative \ch{RE2O2S} powder samples for (b) RE~=~Ce~-~Sm, (c) RE~=~Eu~-~Dy, and d) RE~=~Ho~-~Yb, highlighting the observed magnetic ordering transitions. (e) Expected and calculated effective magnetic moments ($\mu_{\text{eff}}$) and Curie-Weiss temperatures ($\theta_{\text{CW}}$) for each \ch{RE2O2S} phase, derived from high- and low-temperature Curie-Weiss fitting of the measured magnetic susceptibility, respectively.} 
    \label{MvT_main}
\end{figure*}

\begin{figure*}[ht]
    \includegraphics[width=1.0\textwidth]{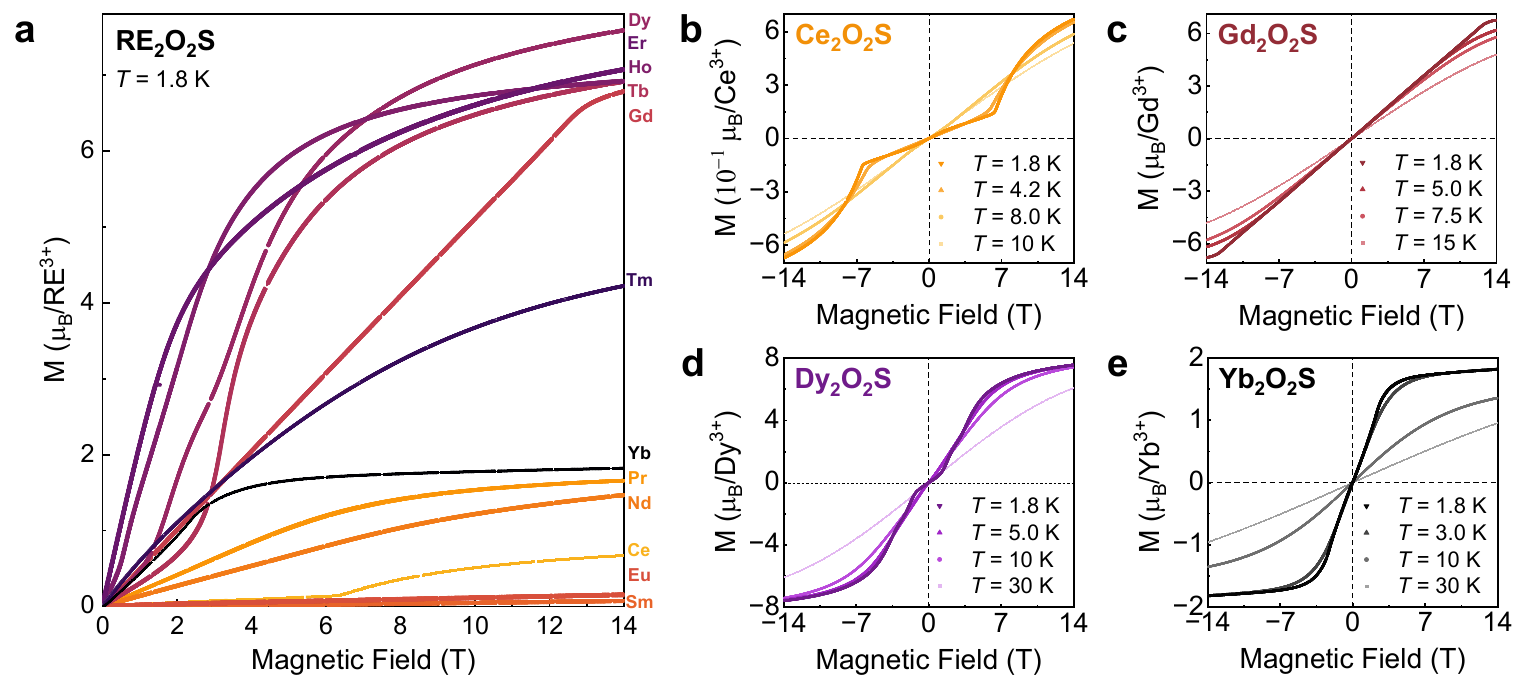}
    \caption{(a) Field-dependent magnetization of representative \ch{RE2O2S} powder samples at \textit{T}~=~1.8~K. Temperature-dependence of the field-dependent magnetization of (b) \ch{Ce2O2S}, (c) \ch{Gd2O2S}, (d) \ch{Dy2O2S}, and (e) \ch{Yb2O2S}.}
    \label{MvH}
\end{figure*}

To explore the low-temperature magnetism of these rare-earth oxysulfides, we performed temperature-dependent DC magnetic susceptibility, $\chi(T)$, and field-dependent DC magnetization, $M(H)$, measurements on all twelve polycrystalline, magnetic \ch{RE2O2S} samples. The supporting data are shown in Figures \ref{MvT_main}, \ref{MvH}, S3, and S4. Curie-Weiss (CW) fitting in both the high- ($T=150$--300~K) and low-temperature range was used to calculate the effective magnetic moment, $\mu_{\text{eff}}$, 
and estimate of net interaction strength, $\theta_{\text{CW}}$, within the paramagnetic regime and close to the ordered state, respectively (Table \ref{CW_fits_full}). The $\theta_{\text{CW}}$ values derived from high-temperature CW fitting in lanthanide compounds are not highlighted here, as they often significantly overestimate the net magnetic interaction strength, owing to contributions from thermally populated CEF levels~\cite{johnston2017influence,mugiraneza2022tutorial}. The frustration index, \textit{f}~=~|$\theta_{\text{CW}}$|/$T_{\text{N}}$, was also calculated for each composition using the $\theta_{\text{CW}}$ derived from the low-temperature Curie-Weiss fits. In cases where no transition was observed, the lowest measured temperature (1.8~K) was used in place of $T_{\text{N}}$ to set a lower bound on $f$. Low-temperature fit-derived $\theta_{\text{CW}}$ values more accurately capture interactions in the magnetic ground state than those derived from higher-temperature fits, where the thermal population of CEF levels dominate. However, as low-temperature fitting can still be influenced by short-range magnetic correlations or the population of low-lying CEF levels, the calculated frustration parameters reported here should be considered only as a means of qualitative comparison between oxysulfide compositions. CEF energy level splittings were calculated from a point charge model using PyCrystalField in the \textit{J} basis for all \ch{RE2O2S} besides RE~=~Eu and Gd, where \textit{J}~=~0 and \textit{L}~=~0 were assumed, respectively \cite{scheie2021pycrystalfield}.

Compared to the more three-dimensional \ch{RE2O3}, magnetic exchange between bilayer slabs in \ch{RE2O2S} is expected to uniformly increase, due to the hybridized, more diffuse electron density on each S$^{2-}$ anion ($J_2$, $J_3$, and $J_5$ in Figure S1). While the highly-localized nature of the 4\textit{f} orbitals on each metal center precludes the clear determination of magnetic exchange strength, this strengthened interslab superexchange and intrinsic structural anisotropy is expected to produce a more anisotropic magnetic response, relative to the fully structurally and magnetically 3D \ch{RE2O3}. \ch{RE2O2S} compositions are thus expected to represent effective intermediates between the purely 3D and 2D analogues (\emph{i.e.} \ch{KREO2}). As a result, intermediate-strength magnetic frustration in these materials is likely to arise from both the geometric frustration intrinsic to the 2D triangular lattice and exchange frustration which dominates in 3D \ch{RE2O3}. 

\noindent\textbf{\ch{Ce2O2S}}: In contrast to earlier reports on \ch{Ce2O2S}, which claimed that it remains paramagnetic down to $T=2$~K~\cite{quezel1970proprietes}, we find that \ch{Ce2O2S} exhibits a sharp cusp at $T_{\text{N}}=8.5$~K in $\chi(T)$, indicative of an antiferromagnetic (AFM) ordering transition (Figure \ref{MvT_main}b, S3a). High-temperature CW fitting yields $\mu_{\text{eff}}=2.39(2)$~$\mu_{\text{B}}$, in good agreement with the expected $\mu_{\text{calc}}=2.54$~$\mu_{\text{B}}$ for Ce$^{3+}$. Fitting at lower temperature ($T=10$--25~K) yields a reduced moment of $\mu_{\text{eff}}=1.99(1)$~$\mu_{\text{B}}$ and $\theta_{\text{CW}}=-9.6(1)$~K. The difference between the low-temperature and high-temperature paramagnetic moments is consistent with the thermal depopulation of excited crystal-field levels and a ground state Kramers doublet with dominant $\ket{m_J}=\ket{\pm \sfrac{3}{2}}$, as supported by the calculated CEF ground state shown in Table \ref{CEF_summary}. Unlike most of the other \ch{RE2O2S} materials studied here, our point charge calculations suggest a well-separated CEF ground state with $\Delta E_{0\rightarrow1}=13.3$~meV. The measured $M(H)$ at \textit{T}~=~1.8~K for \ch{Ce2O2S} displays one sharp inflection around $H_{C}$~=~6.4(1)~T, marking a metamagnetic transition, which gradually broadens upon increasing temperature and is no longer visible at or above $T_{\text{N}}$ (Figure \ref{MvH}b). Up to $H = 14$~T, the magnetization of \ch{Ce2O2S} remains unsaturated, only reaching 0.67~$\mu_{\text{B}}$/Ce$^{3+}$, suggestive of strong anisotropy in its ordered state. \\

\noindent\textbf{\ch{Pr2O2S}}: \ch{Pr2O2S} also exhibits a previously unobserved kink in $\chi(T)$ at $T_{\text{N}}$~=~1.5~K, below which minor splitting between the zero-field-cooled (ZFC) and field-cooled (FC) measurements is observed (Figure \ref{MvT_main}b, S3b). High-temperature CW fitting yields the expected paramagnetic moment, $\mu_{\text{eff}}$~=~3.60(1)~$\mu_{\text{B}}$, for Pr$^{3+}$ ($\mu_{\text{calc}}$~=~3.58~$\mu_{\text{B}}$), while lower-temperature fitting from $T=5$--15~K yields a reduced crystal-field ground state moment of $\mu_{\text{eff}}$~=~3.07(1)~$\mu_{\text{B}}$ and $\theta_{\text{CW}}=-7.1(1)$~K. Our point charge calculation suggests a non-magnetic singlet ground state for Pr$^{3+}$ in \ch{Pr2O2S}, which is possible for a non-Kramers species. However, the calculated CEF splitting between the singlet ground state and pseudo-doublet first excited state ($\Delta E_{0\rightarrow1}=0.89$~meV) is quite small, such that the CEF ground state may function as an effective triplet (Table \ref{CEF_summary}). The measured $M(H)$ for \ch{Pr2O2S} shows no sign of field-induced magnetic phase transitions and only a gradual magnetization plateau upon increasing field strength down to $T=0.4$~K (Figures \ref{MvH}a, S4b). \\

\noindent\textbf{\ch{Nd2O2S}}: \ch{Nd2O2S} displays a very weak cusp in $\chi(T)$ at $T_{\text{N}}$~=~2.2~K, accompanied by minor ZFC-FC splitting upon further cooling (Figure \ref{MvT_main}b, S3c). Although not previously observed down to \textit{T}~=~2.0~K~\cite{quezel1970proprietes,beaury1990interpretation}, the observed ordering transition appears similar to that reported for the structurally-analogous \ch{Nd2O3} at $T_{\text{N}}$~=~0.55~K \cite{rai2020magnetism}. High-temperature CW fitting yields the expected paramagnetic moment, $\mu_{\text{eff}}$~=~3.64(1)~$\mu_{\text{B}}$, which agrees well with $\mu_{\text{calc}}$~=~3.62~$\mu_{\text{B}}$ for Nd$^{3+}$ (Table \ref{CW_fits_full}). Low-temperature (\textit{T}~=~5--15~K) fitting yields a reduced $\mu_{\text{eff}}$~=~3.32(1)~$\mu_{\text{B}}$ and $\theta_{\text{CW}}$~=~-17.6(1)~K, in agreement with a Kramers doublet ground state and weaker frustration than that observed in more three-dimensional \ch{Nd2O3} (Table \ref{CEF_summary}) \cite{rai2020magnetism}. No field-induced features are visible in the $M_{H}$ for \ch{Nd2O2S}, down to \textit{T}~=~1.8~K and up to $H = 14$~T (Figures \ref{MvH}a, S4c).\\

\noindent\textbf{\ch{Sm2O2S}}: A cusp is observed in the measured $\chi(T)$ of \ch{Sm2O2S} at $T_{\text{N}}$~=~3.5~K, below which significant ZFC-FC splitting develops (Figure \ref{MvT_main}b, S3d), in contrast to reports of paramagnetism down to at least \textit{T}~=~2~K \cite{quezel1970proprietes}. High-temperature CW fitting yields a paramagnetic moment of $\mu_{\text{eff}}$~=~1.73(2)~$\mu_{\text{B}}$, consistent with the expected $\mu_{\text{calc}}$~=~1.70~$\mu_{\text{B}}$ for Sm$^{3+}$ (Table \ref{CW_fits_full}). Lower-temperature ($T=5$--15~K) CW fitting yields $\mu_{\text{eff}}$~=~0.80(1)~$\mu_{\text{B}}$, about half of the expected paramagnetic moment and in line with a reduced CEF ground state moment, and $\theta_{\text{CW}}=-16.9(2)$~K, supporting moderate AFM interactions. The measured magnetization of \ch{Sm2O2S} at \textit{T}~=~1.8~K is the weakest of all \ch{RE2O2S} phases, a product of both its highly-reduced moment and its strong single-ion anisotropy, in good agreement with the calculated $g$-tensors (Table \ref{CEF_summary}). It shows no field-induced features observed up to $H = 14$~T (Figures \ref{MvH}a, S4d). \\

\begin{table*}
    \centering
    \caption{Fitting parameters obtained from low- and high-temperature Curie-Weiss fitting of \ch{RE2O2S} powder magnetization measurements. M$_\text{14 T}$ is defined as the magnetization at \textit{T}~=~1.8~K in a $\mu_\text{o}$H~=~14~T applied field. As described in the text, $\theta_{\text{CW}}$ values in the higher temperature range are excluded due to CEF effects.}
    \label{CW_fits_full}
    \begin{tabular}{|c|c|c|c|c|c|c|c|c|c|}
    \hline
     \multicolumn{1}{|c|}{\multirow{2}{*}{\textbf{RE}}} & \multirow{2}{*}{$\bm{J}$} & \textbf{Fit Range} & $\mathbf{T_N}$ & $\bm{\mu_{\text{\textbf{eff}}}}$ & $\bm{\mu_{\text{\textbf{calc}}}}$ & \textbf{M$_{\text{14 T}}$} & $\bm{\theta_{\text{\textbf{CW}}}}$ & \multirow{2}{*}{$\bm{f = \frac{\theta_{\text{CW}}}{T_{\text{N}}}}$} & $\bm{\chi_0}$ \\
\multicolumn{1}{|c|}{} & & (K) & (K) & ($\mu_{\text{B}}$/RE) & ($\mu_{\text{B}}$/RE) & ($\mu_{\text{B}}$/RE) & (K) & &             (emu/mol$_{\text{RE}}$-Oe) \\\hline
    \multirow{2}{*}{Ce} & \multirow{2}{*}{5/2} & 150--300 & \multirow{2}{*}{8.5} &  2.39(2) & \multirow{2}{*}{2.54} & \multirow{2}{*}{0.67} & -38(2) & -- & 0 \\\cline{3-3}\cline{5-5}\cline{8-10}
    & & 10--25 & & 1.99(1) & & & $-9.6(1)$ & 1.1(1) & 0 \\\hline
    \multirow{2}{*}{Pr} & \multirow{2}{*}{4} & 150--300 & \multirow{2}{*}{1.5}  & 3.60(1) & \multirow{2}{*}{3.58} & \multirow{2}{*}{1.66} & -41.0(7) & -- & $-2.8(3) \times 10^{-4}$ \\\cline{3-3}\cline{5-5}\cline{8-10}
    & & 5--15 & & 3.07(1) & & & $-7.1(1)$ & 4.7(1) & $-7.8(1) \times 10^{-3}$ \\\hline
    \multirow{2}{*}{Nd} & \multirow{2}{*}{9/2} & 150--300 & \multirow{2}{*}{2.2}  & 3.64(1) & \multirow{2}{*}{3.62} & \multirow{2}{*}{1.46} & -32.5(6) & -- & 0 \\\cline{3-3}\cline{5-5}\cline{8-10}
    & & 5--15 & & 3.32(1) & & & $-17.6(1)$ & 8.0(1) & 0 \\\hline
    \multirow{2}{*}{Sm} & \multirow{2}{*}{5/2} & 15--300 & \multirow{2}{*}{3.5} & 1.73(2) & \multirow{2}{*}{1.70} & \multirow{2}{*}{0.08} & -208(4) & -- & 0\\\cline{3-3}\cline{5-5}\cline{8-10}
    & & 5--15 & & 0.80(1) & & & $-16.9(2)$ & 4.8(1) & 0 \\\hline
    Gd & 7/2 & 10--300 & 5.6 & 7.95(1) & 7.94 & 6.78 & $-20.1(1)$ & 3.6(1) & $-7.0(3) \times 10^{-4}$\\\hline
    \multirow{2}{*}{Tb} & \multirow{2}{*}{6} & 150--300 & \multirow{2}{*}{9.1} &  9.78(1) & \multirow{2}{*}{9.72} & \multirow{2}{*}{6.91} & -31.9(1) & -- & $-3.0(1)\times10^{-3}$ \\\cline{3-3}\cline{5-5}\cline{8-10}
    & & 25--60 & & 9.20(1) & & & $-16.9(1)$ & 1.9(1) & 0 \\\hline
    \multirow{2}{*}{Dy} & \multirow{2}{*}{15/2} & 150--300 & \multirow{2}{*}{6.1}  & 10.79(1) & \multirow{2}{*}{10.65} & \multirow{2}{*}{7.59} & -33.6(1) & -- & $-3.1(3)\times10^{-3}$\\\cline{3-3}\cline{5-5}\cline{8-10}
    & & 25--60 & & 10.69(1) & & & $-19.3(1)$ & 3.2(1) & $-2.0(1)\times10^{-2}$ \\\hline
    \multirow{2}{*}{Ho} & \multirow{2}{*}{8} & 150--300 & \multirow{2}{*}{2.8}  & 10.54(1) & \multirow{2}{*}{10.61} & \multirow{2}{*}{6.91} & -11.1(1) & -- & 0 \\\cline{3-3}\cline{5-5}\cline{8-10}
    & & 25--60 & & 10.53(1) & & & $-9.7(1)$ & 3.5(1) & 0 \\\hline
    \multirow{2}{*}{Er} & \multirow{2}{*}{15/2} & 150--300 & \multirow{2}{*}{--}  & 9.55(1) & \multirow{2}{*}{9.58} & \multirow{2}{*}{7.07} & -10.5(1) & -- & $-7.0(3)\times10^{-4}$\\\cline{3-3}\cline{5-5}\cline{8-10}
    & & 25--60 & & 9.40(1) & & & $-7.2(1)$ & $>4$ & $-2.5(1)\times10^{-3}$ \\\hline
    Tm & 6 & 150--300 & -- & 7.68(1) & 7.56 & 4.22 & $-24.2(2)$ & -- & $-1.9(3)\times10^{-3}$\\\hline

    \multirow{2}{*}{Yb} & \multirow{2}{*}{7/2} & 150--300 & \multirow{2}{*}{3.0}  & 4.44(1) & \multirow{2}{*}{4.54} & \multirow{2}{*}{1.82} & -49.3(1) & -- & 0 \\\cline{3-3}\cline{5-5}\cline{8-10}
    & & 5--15 & & 3.09(1) & & & $-1.8(1)$ & 0.6(1) & 0 \\\hline
    \end{tabular}
\end{table*}

\noindent\textbf{\ch{Eu2O2S}}: The Eu$^{3+}$ in \ch{Eu2O2S} has a non-magnetic $J=0$ Hund's rules ground state, due to the perfect cancellation of its spin and orbital moments. Consequently, \ch{Eu2O2S} exhibits characteristic van Vleck paramagnetic behavior across the full measured temperature range, owing to thermal population of excited $J$ states (Figure \ref{MvT_main}b, S3e), which is not well captured by a conventional CW fitting. As expected for a non-magnetic ground state, only slight curvature is observed in $M_{H}$ at low fields (Figures \ref{MvH}a, S4e).\\

\noindent\textbf{\ch{Gd2O2S}}: We observe a sharp cusp in the $\chi(T)$ of \ch{Gd2O2S} at $T_{\text{N}}=5.6$~K, in agreement with the literature~\cite{quezel1970proprietes,biondo2014geometric}, with minimal ZFC-FC splitting 
(Figure \ref{MvT_main}b, S3f). Since \textit{L}~=~0 for Gd$^{3+}$, CEF effects are negligible and the CW fitting could be performed over a larger temperature range than for the other compositions. Fitting from $T=10$--300~K yields $\mu_{\text{eff}}$~=~7.95(1)~$\mu_{\text{B}}$ and $\theta_{\text{CW}}=-20.1(1)$~K, in agreement with that expected for Gd$^{3+}$ ($\mu_{\text{calc}}$~=~7.94~$\mu_{\text{B}}$) and supporting the presence of strong AFM exchange interactions (Table \ref{CW_fits_full}). The $M_{H}$ of \ch{Gd2O2S} at \textit{T}~=~1.8~K exhibits no field-induced features before plateauing at 6.78~$\mu_{\text{B}}$/Gd$^{3+}$ above $H = 12$~T, in good agreement with the expected saturated moment of $\mu_{\text{sat}}=g_JJ\mu_{\text{B}}=7$~$\mu_{\text{B}}$ (Figure \ref{MvH}c).\\

\noindent\textbf{\ch{Tb2O2S}}: The $\chi(T)$ of \ch{Tb2O2S} shows a single cusp at $T_{\text{N}}$~=~9.1~K, the highest observed N\'eel ordering transition across the \ch{RE2O2S} family and in agreement with the literature (Figures \ref{MvT_main}c, S3g) \cite{quezel1970proprietes}. High-temperature CW fitting yields $\mu_{\text{eff}}$~=~9.78(1)~$\mu_{\text{B}}$, in good agreement with the expected paramagnetic moment of Tb$^{3+}$, which is only marginally reduced at lower temperature ($T=25$--60~K) to $\mu_{\text{eff}}$~=~9.20(1)~$\mu_{\text{B}}$ (Table~\ref{CW_fits_full}). The low-temperature fit-derived $\theta_{\text{CW}}=-16.9(1)$~K supports the presence of net antiferromagnetic exchange. Point charge calculations support the presence of a doublet ground state, with a low-lying singlet excited state, $\Delta E_{0-1}$~=~0.16 meV (Table \ref{CEF_summary}). The field-dependent magnetization of \ch{Tb2O2S} at $T=1.8$~K displays a single inflection point at around $H = 3.2$~T before plateauing towards its full saturated moment at higher fields (Figures \ref{MvH}a, S4g).\\

\noindent\textbf{\ch{Dy2O2S}}: \ch{Dy2O2S} displays a single cusp at $T_{\text{N}}$~=~6.1~K in $\chi(T)$ (Figures \ref{MvT_main}c, S3h), with high-temperature CW fitting yielding a paramagnetic moment  of $\mu_{\text{eff}}$~=~10.79(1)~$\mu_{\text{B}}$, in good agreement with the Hund's rules moment for Dy$^{3+}$, and no significant reduction at lower temperatures ($T=25$--60~K), where $\theta_{\text{CW}}=-19.3(1)$~K (Table \ref{CW_fits_full}). This low-temperature moment can be attributed to a Kramers doublet of complex character, as determined by point charge calculations presented in (Table \ref{CEF_summary}), as well as the thermal population of a low-lying excited crystal-field state $\Delta E_{0-1}$~=~1.49 meV, whose presence is relevant even at the ordering temperature $T_\text{N} = 6.1$~K. The $M_{H}$ of \ch{Dy2O2S} at $T=1.8$~K displays two clear inflection points, centered around $H = 1.8$~T and $H = 3.6$~T, which are no longer visible by $T=5.0$~K (Figure \ref{MvH}d), pointing to a rich phase diagram for this phase. \\

\noindent\textbf{\ch{Ho2O2S}}: The $\chi(T)$ of \ch{Ho2O2S} shows a single kink at $T_{\text{N}}$~=~2.8~K (Figures \ref{MvT_main}c, S3i), as observed previously \cite{quezel1970proprietes}. The high-temperature CW fitting yields $\mu_{\text{eff}}$~=~10.54(1)~$\mu_{\text{B}}$, in good agreement with the expected paramagnetic moment for Ho$^{3+}$, with no notable reduction in the low-temperature ($T=25$--60~K) limit where $\theta_{\text{CW}}=-9.7(1)$~K (Table \ref{CW_fits_full}). The crystal-field ground state of \ch{Ho2O2S} is formed by a singlet with a low-lying doublet excited state. where $\Delta E_{0-1}$~=~0.22 meV (Table \ref{CEF_summary}). The relatively-low $\theta_{\text{CW}}$ suggests the presence of weaker antiferromagnetic interactions in \ch{Ho2O2S}, compared to most of the the other \ch{RE2O2S} phases reported here. The field-dependent magnetization of \ch{Ho2O2S} at $T=1.8$~K displays two weak inflection points, centered around $H = 0.6$~T and $H = 2.2$~T, before plateauing at higher fields (Figures \ref{MvH}a, S4i).\\

\noindent\textbf{\ch{Er2O2S}}: \ch{Er2O2S} is observed to display paramagnetic behavior in $\chi(T)$ down to \textit{T}~=~1.8~K, just above the onset of short-range order reported at $T = 1.5$~K (Figures \ref{MvT_main}c, S3j) \cite{ballestracci1968etude,ballestracci1969structure}. \ch{Er2O2S} is therefore the only \ch{RE2O2S} phase without observed long-range order, excluding phases with non-magnetic singlet ground states. The high-temperature CW fit yields a paramagnetic moment of $\mu_{\text{eff}}$~=~9.55(1)~$\mu_{\text{B}}$, which again is not significantly reduced at lower-temperatures ($T=25$--60~K). \ch{Er2O2S} also continues the trend started with \ch{Ho2O2S}, showing a significantly reduced $\theta_{\text{CW}}=-7.2(1)$~K (Table \ref{CW_fits_full}), indicating weakened antiferromagnetic exchange interactions. The Kramers doublet ground state for Er$^{3+}$ has a small CEF splitting to the first excited state $\Delta E_{0-1}$~=~2.03 meV (Table \ref{CEF_summary}). As expected for a material in its paramagnetic phase, the $M_{H}$ of \ch{Er2O2S} at $T=1.8$~K exhibits no field-induced features up to $H = 14$~T (Figures \ref{MvH}a, S4j). \\

\begin{table*}
    \centering
    \vspace{-2mm}
    \caption{Calculated ground state multiplet and CEF splitting to the first excited state for anisotropically-coordinated, trivalent rare-earth cations in the \ch{RE2O2S} structure.}
    \label{CEF_summary}
    \renewcommand{\arraystretch}{1.2}
    \begin{tabular}{|c|c|c|c|c|c|c|c|}
    \hline
     \multirow{2}{*}{\textbf{RE}} & \textbf{Ground state} & \multirow{2}{*}{\textbf{Dominant character}} & \textbf{$\Delta$E$_0$~ $\rightarrow$~E$_1$} & \multirow{2}{*}{\textbf{$g_z$}} & \multirow{2}{*}{\textbf{$g_\perp$}} & \textbf{$\mu_z$} & \textbf{$\mu_\perp$} \\
     & \textbf{multiplet} & & (meV) & & & ($\mu_\text{B}$) & ($\mu_\text{B}$) \\\hline
     Ce & Doublet & $|\pm\frac{3}{2}\rangle$ & 13.31 & 2.57 & 0 & 1.29 & 0 \\\hline
     Pr & Singlet & $|\pm3\rangle$, $|0\rangle$ & 0.89 & - & - & - & -\\\hline
     Nd & Doublet & $|\pm\frac{7}{2}\rangle$, $|\pm\frac{5}{2}\rangle$, $|\pm\frac{1}{2}\rangle$ & 5.61 & 0.39 & 3.45 & 0.20 & 1.73 \\\hline
     Sm & Doublet & $|\pm\frac{5}{2}\rangle$, $|\pm\frac{1}{2}\rangle$ & 1.86 & 0.63 & 0.40 & 0.31 & 0.20 \\\hline
     Tb & Doublet & $|\pm5\rangle$, $|\pm4\rangle$, $|\pm2\rangle$, $|\pm1\rangle$ & 0.16 & 1.97 & 0 & 0.99 & 0 \\\hline
     Dy & Doublet & $|\pm\frac{13}{2}\rangle$, $|\pm\frac{11}{2}\rangle$, $|\pm\frac{7}{2}\rangle$, $|\pm\frac{5}{2}\rangle$, $|\pm\frac{1}{2}\rangle$ & 1.49 & 3.14 & 9.62 & 1.57 & 4.81 \\\hline
     Ho & Singlet & $|\pm6\rangle$, $|\pm3\rangle$, $|0\rangle$ & 0.22 & - & - & - & -\\\hline
     Er & Doublet & $|\pm\frac{13}{2}\rangle$, $|\pm\frac{11}{2}\rangle$, $|\pm\frac{7}{2}\rangle$, $|\pm\frac{5}{2}\rangle$, $|\pm\frac{1}{2}\rangle$ & 2.03 & 3.08 & 7.25 & 1.54 & 3.63 \\\hline
     Tm & Singlet & $|\pm6\rangle$, $|\pm3\rangle$ & 0.76 & - & - & - & -\\\hline
     Yb & Doublet & $|\pm\frac{7}{2}\rangle$, $|\pm\frac{3}{2}\rangle$, $|\pm\frac{1}{2}\rangle$ & 32.63 & 3.84 & 3.23 & 1.92 & 1.61 \\\hline
    \end{tabular}
\end{table*}

\noindent\textbf{\ch{Tm2O2S}}: The $\chi(T)$ of \ch{Tm2O2S} shows a largely paramagnetic response, with the exception of a very broad hump centered around $T=10$~K, which was not apparent in the measured $\chi(T)$ for samples containing small \ch{Tm2O3} impurities. This can likely be attributed to thermal depopulation of the CEF into the singlet, non-magnetic ground state, as has been observed in other Tm-based materials (Figures \ref{MvT_main}c, S3k) \cite{awaka2003van}, followed by a weak Curie tail at lower temperature arising from paramagnetic impurities. The calculated crystal-field ground state of Tm$^{3+}$ is a non-magnetic singlet, with a splitting to the first excited state of $\Delta E_{0-1}$~=~0.76 meV (Table \ref{CEF_summary}). We can therefore speculate that the changing curvature in the $\chi(T)$ of \ch{Tm2O2S} is due to the thermal depopulation of these excited crystal-field states as Tm$^{3+}$ effectively loses its moment. High-temperature CW fitting yields $\mu_{\text{eff}}$~=~7.68(1)~$\mu_{\text{B}}$, which agrees well with the expected paramagnetic value (Table \ref{CW_fits_full}), however lower-temperature fitting could not be reliably performed. The field-dependent magnetization of \ch{Tm2O2S} at $T=1.8$~K displays only a gradual increase up to $H = 14$~T  (Figures \ref{MvH}a, S4k). The increasing moment can be explained by Zeeman splitting of the CEF states, changing the population of the excited CEF levels and thereby enhancing the paramagnetic moment.\\

\noindent\textbf{\ch{Yb2O2S}}: \ch{Yb2O2S} is observed to display a sharp cusp in $\chi(T)$ at $T_{\text{N}}$~=~3.0~K, in agreement with Quezel \textit{et. al.} ((Figures \ref{MvT_main}c, S3l)) \cite{quezel1970proprietes}. High-temperature CW fitting yields $\mu_{\text{eff}}$~=~4.44(1)~$\mu_{\text{B}}$, in agreement with the $\mu_{\text{calc}}$~=~4.54~$\mu_{\text{B}}$ expected for Yb$^{3+}$ (Table \ref{CW_fits_full}). However, low-temperature ($T=5$--15~K) fitting yields a reduced moment of $\mu_{\text{eff}}$~=~3.09(1)~$\mu_{\text{B}}$ and a very low $\theta_{\text{CW}}$~=~-~1.8(1)~K, which is lower than $T_{\text{N}}$ and a continuation of the trend that started with \ch{Ho2O2S} of decreasing CW temperatures. \ch{Yb2O2S} is notable for having a large CEF splitting, $\Delta E_{0-1}$~=~32.6 meV (Table \ref{CEF_summary}). The $M_{H}$ of \ch{Yb2O2S} at \textit{T}~=~1.8~K displays two field-induced features at $H = 0.2$~T and $H = 2.3$~T, before plateauing at relatively low fields (Figure \ref{MvH}e).\\

The variability in the observed magnetic response across the \ch{RE2O2S} series indicates a diversity of largely AFM-ordered ground states and field-induced metamagnetic phase transitions. This is also reflected in our CEF calculations, which suggest a wide range of ground state degeneracies, ranging from non-magnetic singlets (RE~=~Pr, Ho, and Tm) to well-separated doublets and effective triplets, as well as three major classes of single-ion anisotropies - Ising-like for RE~=~Ce and Tb, XY-like for RE~=~Nd, and more isotropic for RE~=~Sm, Dy, Er, and Yb (Table \ref{CEF_summary}). \ch{Ce2O2S} is particularly notable, as its predominantly m$_J$~=~$\pm\frac{3}{2}$ ground state doublet has dipolar-octupolar character. The transitions suggestive of long-range order at low temperatures for both \ch{Pr2O2S} and \ch{Ho2O2S} additionally highlights the relevance of ground state mixing with low-lying excited states. Understanding the nature of magnetic ground states across the \ch{RE2O2S} family will then require more thorough consideration of single-ion physics.\\


The observation of previously unreported transitions consistent with long-range AFM ordering for RE~=~Ce, Pr, Nd, and Sm, as well as sensible experimentally derived paramagnetic moments for all compositions, suggests a higher degree of structural order in \ch{RE2O2S} samples grown by the method described in this work. Both Ce and Nd compositions are observed to order higher in temperature and with smaller calculated frustration parameters (using either low- or high-temperature-derived $\theta_{\text{CW}}$) than in the isostructural, more three-dimensional \ch{Ce2O3} ($T_{\text{N}}$~=~6.2~K, \textit{f}~=~9.2(1)~\cite{cote2023direct}) and \ch{Nd2O3} ($T_{\text{N}}$~=~0.55~K, \textit{f}~=~18.5(1)~\cite{rai2020magnetism}). This trend appears to originate from the overall larger distances between RE$^{3+}$ moments as well as the greater difference between RE$^{3+}$ distances within and across slabs, effectively reducing the degeneracy of equivalent interactions. Interestingly, \textit{f} is smallest for \ch{Ce2O2S} and \ch{Yb2O2S}, both of which have the largest CEF splitting to the first excited state and would thus be expected to possess less classical, more reliably $S_{\text{eff}}$~=~1/2 ground states.

Despite the broad air stability of these materials described in previous reports and negligible changes observed in the measured pXRD over time, nearly all compositions showed significant changes in the measured $\chi(T)$ upon prolonged exposure to air. This magnetic disorder typically manifested as a broadening of the observed magnetic ordering transition, greater ZFC-FC splitting, and larger calculated effective magnetic moments and Curie-Weiss temperatures. With the exception of \ch{Ce2O2S}, where irreversible oxidation to the mixed-valent \ch{Ce2O_{2.5}S} was observed beyond 30 minutes of air exposure \cite{mauricot1997crystal}, re-annealing of each sample with excess sulfur was found to gradually recover the magnetic response of the as-synthesized material, suggesting that air exposure promotes the loss of sulfur from the \ch{RE2O2S} structure.

\subsection{Specific heat}

\begin{figure}[ht]
    \includegraphics[width=0.40\textwidth]{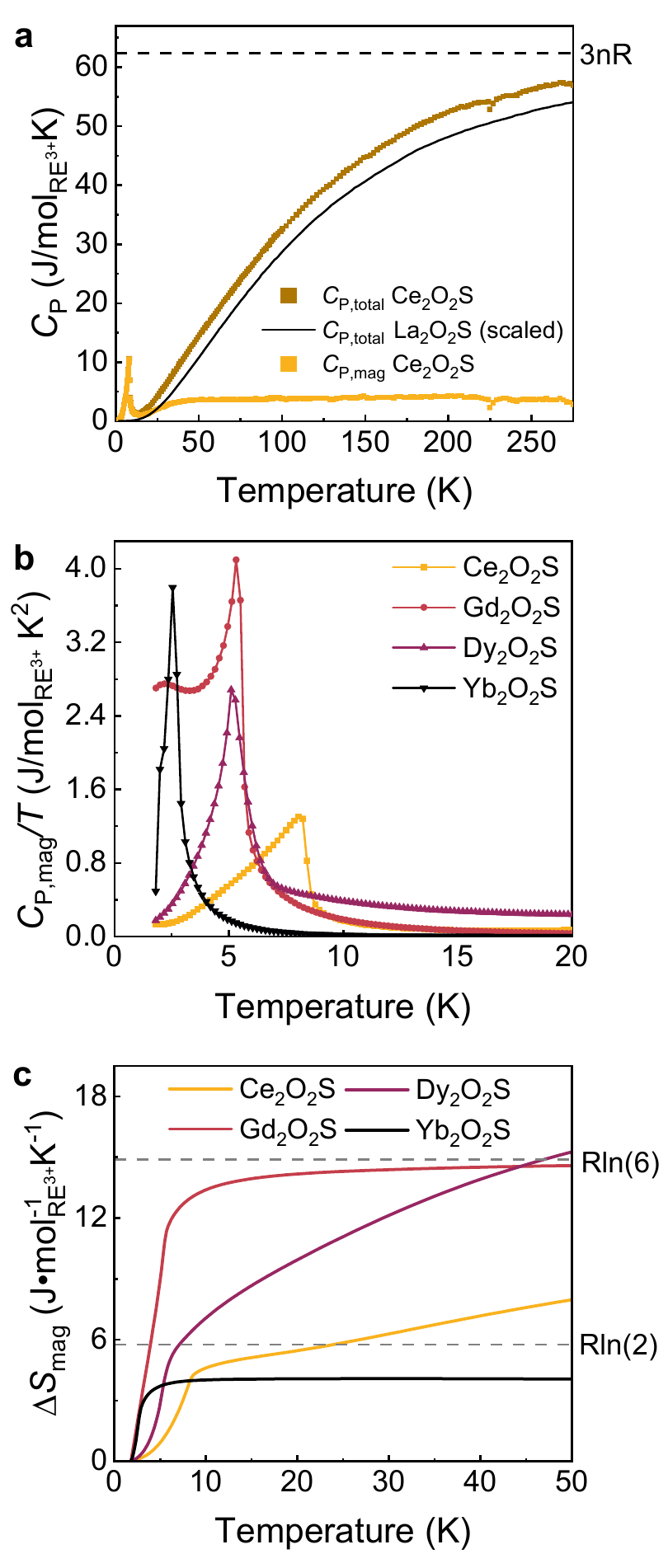}
    \caption{a) Zero-field specific heat as a function of temperature for \ch{Ce2O2S} (brown), \ch{La2O2S} (black line), and their difference (gold). (b) Zero-field magnetic specific heat divided by temperature as a function of temperature for \ch{Ce2O2S} (gold), \ch{Gd2O2S} (red), \ch{Dy2O2S} (purple), and \ch{Yb2O2S} (black) powder samples. (c) Integrated magnetic entropy rise as a function of temperature for each composition up to \textit{T}~=~50~K.}
    \label{HC_main}
\end{figure}

To further investigate the magnetic ground states of the rare-earth oxysulfides, \ch{RE2O2S},
specific heat as a function of temperature was measured for several compositions (RE~=~Ce, Gd, Dy, Yb). The selected compositions were chosen to represent a range of anisotropies and, by extension, likely magnetic ground states. In each case, the specific heat was measured from \textit{T}~=~2--300~K and observed to reach a value close to the Dulong Petit limit ($3nR$) close to room temperature. The magnetic contribution to the specific heat was isolated through the subtraction of an appropriate lattice standard, either \ch{La2O2S} or \ch{Lu2O2S} for the early and latter rare earths, respectively, which was scaled by a correction for the molar mass, as shown for \ch{Ce2O2S} in Figure \ref{HC_main}a. Numerical integration of the low-temperature magnetic heat capacity, $C_{\text{P,mag}}/T$, was used to determine the entropy release, $\Delta S_{\text{mag}}$ associated with the magnetic ordering transition. \\

\noindent\textbf{\ch{Ce2O2S}}:
The specific heat of \ch{Ce2O2S}, shown in Figure \ref{HC_main}a, shows a single lambda-like anomaly at $T_\text{N}=8.5$~K, in agreement with the ordering transition observed in $\chi(T)$. Scaled subtraction of the measured specific heat of the La-analogue and subsequent integration of $C_{\text{P,mag}}/T$ yields an initial rise in magnetic entropy, $\Delta S_{\text{mag}}$, of roughly 80\% of the $R\ln{(2)}$ associated with a ground state doublet up to \textit{T}~=~10~K, and reaches 100\% just above 20~K (Figure \ref{HC_main}b,c). This suggests that significant short-range correlations may develop well above $T_N$, as is often observed in frustrated antiferromagnets.  A gradual increase is then observed up to $\Delta S_{\text{mag}}$~=~14.81~$\text{J}\cdot \text{mol}_{\text{Ce}^{3+}}^{-1}\text{K}^{-1}$, close to the $R\ln{(6)}=14.90$ $\text{J}\cdot \text{mol}^{-1}\text{K}^{-1}$ expected for a $J= 5/2$ system by $T=300$~K, and supporting the steady population of CEF levels up to ambient temperature (Figure S5e). \\

\noindent\textbf{\ch{Gd2O2S}}:
The specific heat of \ch{Gd2O2S} is highly distinctive, containing a sharp anomaly at $T_{\text{N}}=5.6$~K, followed by a second, broader anomaly at $T=2.2(2)~$K \cite{rossat1974crystal3}. Close comparison with the susceptibility data shows that this second anomaly is coincident with a minor slope change in $\chi(T)$ (Figure \ref{HC_main}b). This additional feature was reproducible across \ch{Gd2O2S} sample batches and measurements, suggesting the presence of an intrinsic lower-temperature moment reorientation below $T_{\text{N}}$. Despite the largely linear shift observed in vibrational frequencies from RE~=~La~-~Lu, the scaled $C_{\text{P,total}}$ for \ch{La2O2S} yielded convergence with the Gd curve before surpassing it above \textit{T}~=~50~K, whereas the scaled Lu dataset failed to converge with the Gd analogue up to \textit{T}~=~300~K. This could be due to larger changes in O/S ordering between even the best-ordered samples of each composition. A weighted average of the two scaled $C_{\text{P,total}}$ for \ch{La2O2S} and \ch{Lu2O2S}, 44~La~:~56~Lu, was ultimately found to best approximate the phononic contribution to the measured $C_{\text{P,total}}$ of \ch{Gd2O2S}. By \textit{T}~=~50~K, $\Delta S_{\text{mag}}$ for this system peaks at 14.59~$\text{J}\cdot \text{mol}_{\text{Gd}^{3+}}^{-1}\text{K}^{-1}$, much lower than the Rln(8) expected for Gd$^{3+}$ with \textit{S}~=~7/2, before plateauing up to higher temperatures (Figures \ref{HC_main}c, S5e). As the lower-temperature anomaly remains far from reaching zero down to \textit{T}~=~1.8~K, the significant remaining entropy must be released below it.\\

\begin{figure*}[ht]
    \includegraphics[width=0.8\textwidth]{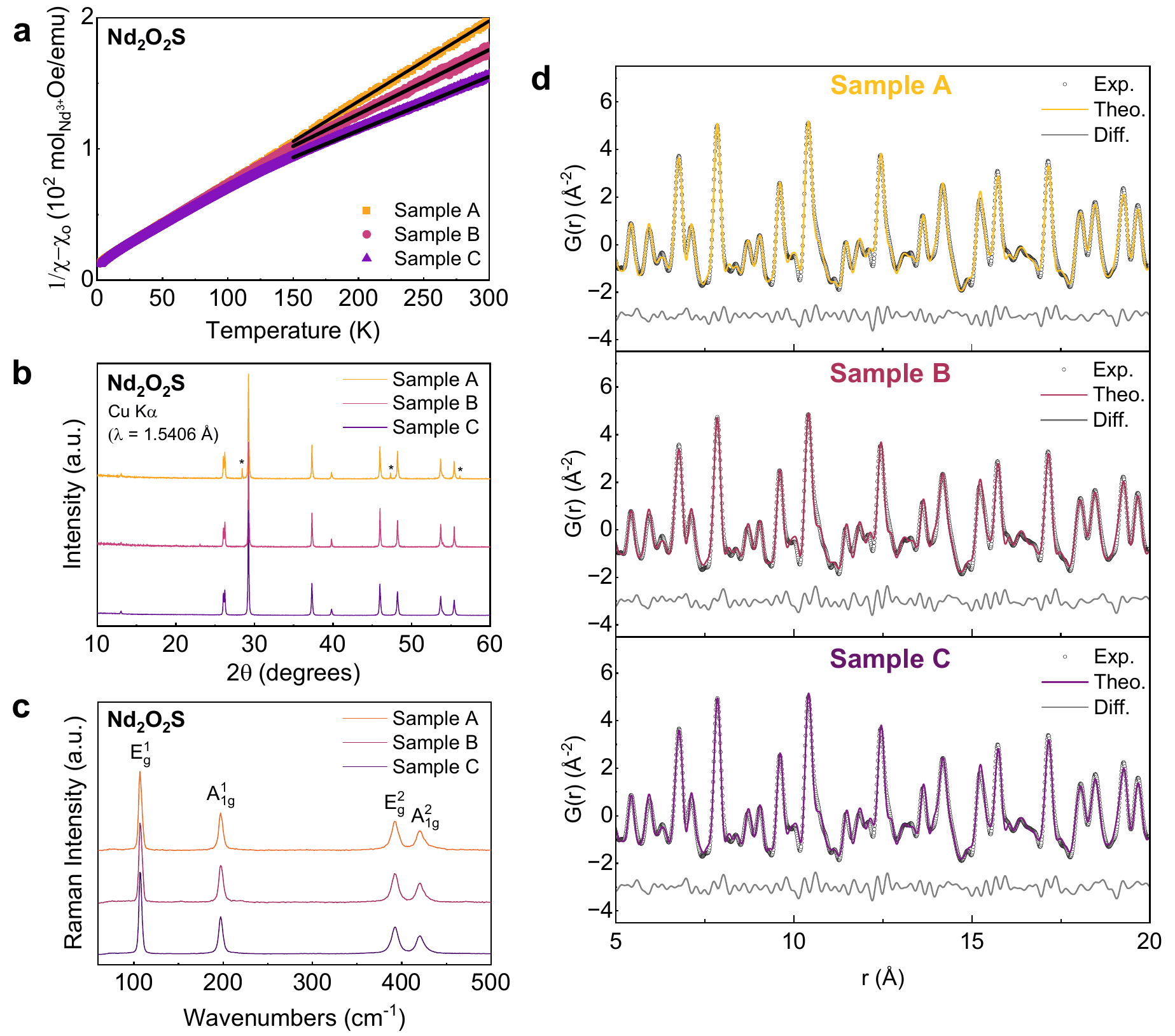}
    \caption{(a) Inverse magnetic susceptibility as a function of temperature, (b) powder diffraction patterns, and (c) Raman scattering spectra collected from \ch{Nd2O2S} powder samples A (gold), B (magenta), and C (purple). The stars in (a) denote peaks attributed to an internal Si standard and the solid black lines in (c) represent the high-temperature Curie-Weiss fit for each sample. (d) Reduced pair distribution function [G(r)] collected from each \ch{Nd2O2S} powder sample. Experimental datasets are shown in black, theoretical fits are colored by sample as in (a)-(c), and the difference curve between them is shown in gray.}
    \label{PDF_fig}
\end{figure*}

\noindent\textbf{\ch{Dy2O2S}}:
The specific heat of \ch{Dy2O2S}, shown in Figure \ref{HC_main}b, shows a well defined lambda anomaly centered around $T=5.1(2)$~K, in agreement with the ordering transition observed in $\chi(T)$ (Figure \ref{HC_main}b). Above the ordering anomaly, we can notice significant additional entropy release up to \textit{T}~=~20 K by comparison with the other measured \ch{RE2O2S} samples, which we can attribute to the presence of a low-lying first excited CEF level ($\Delta E_{0\rightarrow1}=1.49$~meV. Similar to the Gd member, the scaled subtraction of the $C_{\text{P,total}}$ for either \ch{La2O2S} and \ch{Lu2O2S} alone failed to yield a sensible trend in $\Delta S_{\text{mag}}$ for the Dy member. A weighted average of the two scaled datasets, 35~La~:~65~Lu, was ultimately found to reasonably approximate the phononic contribution to $C_{\text{P,total}}$. Although the increase in $\Delta S_{\text{mag}}$ is gradual over the measured temperature range, a slope change is observed around $T=7$~K, just above $T_{\text{N}}$ for the system and coinciding with Rln(2), the expected $\Delta S_{\text{mag}}$ for a ground state doublet. $\Delta S_{\text{mag}}$ gradually rises over the measured temperature range, consistent with the steady population of CEF levels, before saturating around 21.9~$\text{J}\cdot \text{mol}_{\text{Dy}^{3+}}^{-1}\text{K}^{-1}$ at \textit{T}~=~300~K, close to the 23.1~$\text{J}\cdot \text{mol}^{-1}\text{K}^{-1}$ expected for a $J=15/2$ system (Figure S5e).\\

\noindent\textbf{\ch{Yb2O2S}}:
The specific heat of \ch{Yb2O2S} shows only one sharp lambda anomaly at \textit{T}~=~2.5~K, close to the $\chi(T)$-derived $T_{\text{N}}=3.0$~K (Figure \ref{HC_main}b). Scaling of the $C_{\text{P,total}}$ for \ch{Lu2O2S} provides nearly exact agreement between the datasets from $T=15$--55~K, below which integration yields $\Delta S_{\text{mag}}$~=~4.06~$J\cdot \text{mol}_{\text{Yb}^{3+}}^{-1}\text{K}^{-1}$ (Figure \ref{HC_main}c). This value is significantly lower than the $R\ln{(2)}$ for a $S_{\text{eff}}$~=~1/2 Kramers doublet ground state, corresponding to only two-thirds of the expected entropy release. This suggests that there may be additional entropy release below the measured temperature range or that there is a dynamic component to the state realized below $T_{\text{N}}=3.0$~K. The negligible $C_{\text{mag}}$ in the intermediate temperature regime nicely reflects the large CEF splitting to the first excited state highlighted in Table \ref{CEF_summary}. Above this region, $\Delta S_{\text{mag}}$ gradually increases to 9.40~$J\cdot \text{mol}_{\text{Yb}^{3+}}^{-1}\text{K}^{-1}$, well below the expected Rln(8)~=~17.28~$J\cdot \text{mol}^{-1}\text{K}^{-1}$ for the full \textit{S}~=~7/2 Yb$^{3+}$ moment (Figure S5e).

\subsection{Strong disorder-dependence}
In the course of this study, we observed significant and previously-unreported sample dependence in the measured magnetization of each oxysulfide material. This sample dependence manifests as deviations from the expected paramagnetic moment, $\mu_{\text{eff}}$, and variations in the CW temperature, $\theta_{\text{CW}}$, as determined by Curie-Weiss fits in the high-temperature limit. To identify the origin of this sample dependence, we grew the Nd member by three variations on the standard solid-state method. The first proceeded by the annealing of stoichiometric quantities of each precursor, followed by subsequent regrindings and reannealings until \ch{Nd2O3} impurities no longer appeared in the measured pXRD pattern (hereafter referred to as Sample B). The second continued in the same manner with a 5\% molar excess of sulfur added between each annealing until the calculated effective magnetic moment, derived from high-temperature CW fitting of the measured magnetic susceptibility, approached the expected 3.62 $\mu_{\text{B}}$/Nd$^{3+}$ (Sample A). The third method proceeded with the inclusion of a 0.5 mmol chunk of aluminum metal in an alumina crucible sealed above the other reagents in the same quartz tube (Sample C). Here, the aluminum acted as both an oxygen- and sulfur-getter.

Pronounced differences are observed in the measured magnetic susceptibility of each sample, shown in Figure \ref{PDF_fig}a. Both Sample A and Sample B display a slight kink at $T_{\text{N}}$~=~2.3~K and no ZFC-FC splitting, whereas Sample C exhibits a broader kink centered at $T_{\text{N}}$~=~2.2~K and ZFC-FC splitting up to \textit{T}~=~150 K. Curie-Weiss fitting of the inverse susceptibility from \textit{T}~=~150~-~300~K reveals a gradual increase in the degree of deviation from Curie-Weiss behavior with presumed increasing disorder. The calculated effective magnetic moment and Curie-Weiss temperature for Sample A are largely in agreement with literature values [$\mu_{\text{eff}}$~=~3.62(1)~$\mu_{\text{B/Nd}^{3+}}$, $\theta_{\text{CW}}$~=~-22.8(6)~K], while both values increase for Sample B [$\mu_{\text{eff}}$~=~4.04(1)~$\mu_{\text{B/Nd}^{3+}}$, $\theta_{\text{CW}}$~=~-58.6(6)~K] and Sample C [$\mu_{\text{eff}}$~=~4.41(1)~$\mu_{\text{B/Nd}^{3+}}$, $\theta_{\text{CW}}$~=~-77.6(6)~K] (Table \ref{Oxysulfide_CW_fits}). Fitting at low temperatures (\textit{T}~=~5--15~K) instead yields good agreement between Sample A ($\mu_{\text{eff}}$~=~3.41(1)~$\mu_{\text{B/Nd}^{3+}}$ and $\theta_{\text{CW}}$~=~-17.3(1)~K) and the more disordered compositions, suggesting that a larger effective paramagnetic moment does not necessarily indicate significant changes to the low-temperature magnetism. 
As was observed across all \ch{RE2O2S} compositions, the further annealing of Sample B with additional sulfur results in both the expected $\mu_{\text{eff}}$ for Nd$^{3+}$ and a more sensible $\theta_{\text{CW}}$ for a low-temperature, rare-earth antiferromagnet. 

In contrast, room-temperature pXRD patterns measured from each sample, self-normalized to the maximum intensity and shown in Figure \ref{PDF_fig}b, appear nearly identical, with only a slight broadening of reflections observed for Sample C, relative to Sample A and Sample B. The preferred orientation observed across the \ch{RE2O2S} series also presents itself across all three samples, with Rietveld refinement yielding comparable lattice parameters and isotropic displacement parameters. Representative room-temperature Raman scattering spectra collected from each sample, which should be highly-sensitive to the wider distribution of bond lengths in a more disordered material, also appear identical across the three spectra, with the exception of a small hump attributed to a minimal \ch{Nd2O3} impurity in Sample A (Figure \ref{PDF_fig}c). 

This minimal variation in bands arising from Raman-active Nd-O vibrations suggests that sulfur content plays a significant role in determining the physical properties of \ch{RE2O2S} materials, likely through intrinsic sulfur deficiencies, sulfur-oxygen site mixing, or Nd$^{3+}$~-~S$^{2-}$ Schottky defects \cite{suponitskii1988lanthanide}. Interestingly, relative to Sample A, the calculated $\mu_{\text{eff}}$ for Samples B and C are larger than could be attributed to changes in the effective molar mass by any of these disordering mechanisms while maintaining overall structural stability, suggesting that some additional mechanism must be considered. To further probe the structural origin of the magnetic disorder in this system, synchrotron pair distribution function (PDF) analysis was performed on the three representative samples. The reduced PDF [G(r)] data shown in Figure \ref{PDF_fig}d, depicting the probability of finding an atom some distance \textit{r} from another in the \ch{Nd2O2S} structure, appears quite similar between samples, similar to the pXRD data. Refinement of the three datasets against the average \ch{Nd2O2S} structure yields nearly identical lattice parameters and atomic positions for each, within uncertainty (Table \ref{PDF_parameters}). However, refinement of both the oxygen and sulfur site occupancies suggests negligible sulfur incorporation on the oxygen site, in agreement with the uniformity of our Nd-O vibrational spectra, and significant oxygen substitution on the sulfur site coincident with the gradual increase in magnetic disorder (Figure S6). Sample A is found to possess up to 37~$\pm$~16\% oxygen substitution for sulfur on the interslab anion site, greater than previously reported for \ch{La2O2S} (\cite{Borodulenko1985}), while Sample B and Sample C appear even more oxygen-rich with 58~$\pm$~17\% and  78~$\pm$~15\% substitution, respectively. Although likely exaggerated relative to the true level of substitutional disorder, these values are in general agreement with the previous assumption of structural instability for a fully stoichiometric \ch{RE2O2S} phase \cite{suponitskii1988lanthanide}. Rare-earth oxysulfides possessing the triangular-bilayer slab geometry can then be best described by \ch{RE2O_{2+x}S_{1-x}}, where the trigonal structure is stable for all but \textit{x}~$\leq$~0.10 under standard synthesis conditions. 

Attempts to quantitatively assess the compositional differences between the three \ch{Nd2O2S} samples via energy-dispersive X-ray spectroscopy (EDS) measurements resulted in significant beam-induced decomposition. Similar effects, including the removal of sulfur from the sample surface under vacuum conditions, have been previously observed in several studies of rare-earth oxysulfides, even at beam energies as low as 1 keV \cite{holloway2000degradation,jaffar2019cathodoluminescence}. As such, future studies would benefit from the use of x-ray absorption spectroscopy techniques to probe the true extent of disorder on the sulfur sublattice at the sulfur K-edge and accessible RE edges.

The stability of the triangular bilayer slab geometry across nearly the full range of \ch{Nd2O_{2+x}S_{1-x}} stoichiometries naturally raises the question of compositional variability in these and other \ch{RE2O2S} samples synthesized by different methods \cite{larquet2020metal}. Rare-earth oxysulfides have historically been grown by a wide variety of techniques, including by more diffusion-limited solid-state methods than those reported in this work \cite{ballestracci1968etude,zachariasen1949crystal}, or by more disorder-prone solution- and gas-phase decomposition and sulfurization processes \cite{le2025energy,biondo2014geometric,quezel1970proprietes,eick1958preparation}. However, despite this thermodynamic tendency toward structural disorder, studies have thus far largely focused only on its impact on luminescent properties \cite{mikami1999first}. With this in mind, we can reexamine the new magnetic transitions observed in our compositions. In the case of RE~=~Pr and Nd, both transitions are fairly weak and occur close to the base temperature of most commercial cryostats. In previously measured samples, which may have had more substantial disorder, these transitions could have been suppressed below the window of measurement. By contrast, the higher-temperature, more pronounced transitions and increased air-sensitivity observed for our RE~=~Ce and Sm samples can likely be attributed to greater deviations from the ideal stoichiometry in previous reports, due to valence instability. Ce$^{3+}$ and Sm$^{3+}$ are particularly prone to oxidation and reduction to non-magnetic Ce$^{4+}$ and Sm$^{2+}$, respectively, both of which could dilute the magnetic sublattice enough to destabilize long-range magnetic order.

Compared to the oxysulfides, other \ch{RE2O2X} phases are generally expected to be less prone to substitutional disorder on the X sublattice, due to the greater difference in ionic radius and polarizability between oxygen and the heavier chalcogenides. Where more similar, such as in the rare-earth oxyfluorides (REOF), significant synthesis-dependence has been observed in both overall structural stability and defect formation \cite{wen2017polymorphism}. Beyond monatomic anions, even more pronounced interslab disorder has been introduced via the introduction of longer, more flexible polyanionic species, such as cyanamide and carbonate in the isostructural \ch{RE2O2CN2} and \ch{RE2O2CO3} families, respectively \cite{srinivasan2005synthesis,turcotte1969rare}. When considered alongside the broader class of mixed-anion materials, this work serves as an additional reminder of the importance of synthesis technique and local probe structural characterization in the exploration of new heteroanionic magnetic materials.

\begin{table}
    \centering
    \caption{Fitting parameters obtained from Curie-Weiss analysis of \ch{Nd2O2S} powder magnetization measurements.}
    \label{Oxysulfide_CW_fits}
    \begin{tabular}{|c|c|c|c|c|c|}
    \hline
     \multirow{2}{*}{\textbf{Sample}} & \textbf{\emph{T}$_\text{N}$}  & \textbf{Fit Range} & \textbf{$\mu_{\text{eff}}$}  & \textbf{$\theta_{\text{CW}}$} & \multirow{2}{*}{$\chi_\text{o}$ ($10^{-4}$)}\\ 
    & (K) & (K) & ($\mu_\text{B}$/Nd$^{3+}$) & (K) & \\\hline
    A & 2.3  & 150--300 & 3.62(1) & $-22.8(6)$ & 0 \\\hline
    B & 2.3  & 150--300 & 4.04(1) & $-58.6(6)$ & $-8.2(3)$\\\hline
    C & 2.2 & 150--300 & 4.41(1) &$ -77.6(6)$ & 0 \\
  \hline
    \end{tabular}
\end{table}

\begin{table}
    \centering
    \caption{Fitting parameters obtained from PDF analysis of \ch{Nd2O2S} powder samples.}
    \label{PDF_parameters}
    \begin{tabular}{|c|c|c|c|c|c|}
    \hline
     \multicolumn{3}{|c|}{} & \textbf{Sample A} & \textbf{Sample B} & \textbf{Sample C} \\\hline
     \multicolumn{2}{|c|}{} & \textit{a} (\AA) & 3.944(2) & 3.944(2) & 3.945(2) \\\hline
     \multicolumn{2}{|c|}{} & \textit{c} (\AA) & 6.788(4) & 6.787(5) & 6.787(5) \\\hline
     \multirow{2}{*}{Nd1} & & \textit{z} & 0.2803(6) & 0.2800(6) & 0.2802(6) \\
     \multirow{2}{*}{(\textit{2d})} & $\frac{1}{3}$, $\frac{2}{3}$, z & $U_{11}$ & 0.0035(2) & 0.0033(2) & 0.0031(2) \\
      & & $U_{33}$ & 0.0028(5) & 0.0026(5) & 0.0026(4) \\\hline
     & & \textit{z} & 0.632(6) & 0.633(5) & 0.633(5) \\
     \multirow{2}{*}{O1} & & $U_{11}$ & 0.013(4) & 0.014(4) & 0.018(5) \\
     \multirow{2}{*}{(\textit{2d})} & $\frac{1}{3}$, $\frac{2}{3}$, z & $U_{33}$ & 0.006(7) & 0.002(5) & 0.007(4)  \\
     & & O occ. & 0.99(8) & 0.98(8) & 0.99(8) \\
     & & S occ. & 0.01(8) & 0.02(8) & 0.01(8) \\\hline
      & & $U_{11}$ & 0.007(2) & 0.008(3) & 0.005(2) \\
     S1 & \multirow{2}{*}{0, 0, 0} & $U_{33}$ & 0.004(4) & 0.002(4) & 0.001(3) \\
     (\textit{1a}) & & S occ. & 0.63(16) & 0.42(17) & 0.22(15) \\
     & & O occ. & 0.37(16) & 0.58(17) & 0.78(15) \\\hline
     
    \end{tabular}
\end{table}

\section{Conclusion}

In conclusion, we present the synthesis and study of a series of magnetically-frustrated, mixed-anion rare-earth oxysulfides. The unique bilayer slab lattice geometry serves as a structural intermediate between the purely 2D triangular monolayer and 3D \ch{La2O3}-structure type, giving rise to quasi-2D magnetism and intermediate frustration. Magnetic susceptibility measurements reveal several previously-unobserved transitions consistent with long-range AFM order at low-temperatures, as well as a variety of field-induced metamagnetic transitions. Across the \ch{RE2O2S} series, a diversity of magnetic ground states are observed, owing to strong single-ion anisotropy in this highly anisotropic structure. Zero-field specific heat measurements and point charge calculations provide additional insight into the unique CEF excitation spectrum produced by the anisotropic coordination of the rare-earth cation, with the Gd and Yb compositions being particularly worthy of further study, due to the significant magnetic entropy remaining below \textit{T}~=~1.8~K. We also establish a tendency towards sample dependence across the \ch{RE2O2S} series, which we attribute to substitutional anionic disorder on the sulfur site, even for samples displaying minimal signs of magnetic disorder. This work demonstrates the broad tunability of magnetic states enabled by the introduction of mixed-anion character into rare-earth AFMs, as well as the importance of often overlooked structural disorder in the synthesis of new magnetic materials.

\section*{Supporting Information}

Rietveld refinement tables, Raman scattering spectra, and additional magnetic susceptibility, magnetization, specific heat, and simulated reduced pair distribution function [G(r)] data.

\section*{Acknowledgments}

The authors thank J. Dadap and Z. Ye for technical assistance in the collection of the Raman scattering spectra. This work was supported by the Natural Sciences and Engineering Research Council of Canada (NSERC), the Canadian Institute for Advanced Research (CIFAR), the Sloan Research Fellowships program, and the Killam Accelerator Research Fellowship. Part or all of the research described in this paper was performed at the Canadian Light Source, a national research facility of the University of Saskatchewan, which is supported by the Canada Foundation for Innovation (CFI), the Natural Sciences and Engineering Research Council (NSERC), the Canadian Institutes of Health Research (CIHR), the Government of Saskatchewan, and the University of Saskatchewan. The identification of any commercial product or trade name does not imply endorsement or recommendation by the National Institute of Standards and Technology. \\

\bibliography{Refs.bib}


\clearpage

\renewcommand{\thefigure}{S\arabic{figure}}
\renewcommand{\thetable}{S\arabic{table}}
\renewcommand{\theequation}{S\arabic{equation}}
\renewcommand{\thepage}{S\arabic{page}}
\setcounter{figure}{0}
\setcounter{table}{0}
\setcounter{equation}{0}
\setcounter{page}{1} 

\title{Supplementary Materials for:\\ Rare-earth oxysulfides \ch{RE2O2S} as model mixed-anion frustrated magnets}
\maketitle

\section*{Methods}

\textbf{Powder synthesis.} Polycrystalline \ch{RE2O2S} samples were grown by the solid state synthesis method. RE metal, RE oxide, and sulfur (as S~-~Puratronic, 99.999\%, repurified by low-temperature annealing with charcoal at 100°C)) were combined in a 2:2:3 (RE = La, Nd-Gd, Ho-Lu), 1:1:1 (RE = Ce), or 6:2:7 (RE = Tb) molar ratio, for a total input mass of 1.0g, and sealed under vacuum in a 14 mm ID x 16 mm OD x 10 cm long quartz ampoule (Table \ref{Reagents}). For RE = La - Er, the samples were heated to 250°C over 5 hours, held for 5 hours, heated to 500°C over 5 hours, held for 5 hours, heated to 900°C over 5 hours, and held for 48 hours before cooling to room temperature at 100°C per hour. Samples were then reground in an argon-filled glovebox, sealed in new quartz tubes under vacuum, and reheated by the same profile until rare-earth oxide impurity phases were no longer visible in the measured powder X-ray diffraction patterns. Growth of the RE = Tm - Lu and Y phases proceeded by the same method, albeit with a maximum dwell temperature of 1100°C and slower cooling rate of 50°C per hour to force sulfur incorporation.

\begin{table*}
    \centering
    \vspace{-3mm}
    \caption{Reagents used to synthesize representative polycrystalline \ch{RE2O2S} samples. *\ch{Tm2O3} was produced from Tm metal (below) annealed at 1000°C.}
    \label{Reagents}
    \begin{tabular}{|c|c|c|c|}
    \hline
    \textbf{RE metal} & \textbf{RE metal source} & \textbf{RE oxide} & \textbf{RE oxide source}\\\hline
    La & Smart Elements, 99.9\% & \ch{La2O3} & Alfa Aesar, Reacton, 99.99\% \\\hline
    Ce & Smart Elements, 99.9\% & \ch{CeO2} & Alfa Aesar, Reacton, 99.99\% \\\hline
    Pr & Smart Elements, 99.9\% & \ch{Pr6O11} & Alfa Aesar, Reacton, 99.9\% \\\hline
    Nd & Smart Elements, 99.9\% & \ch{Nd2O3} & Thermo Scientific, 99.9\% \\\hline
    Sm & Smart Elements, 99.99\% & \ch{Sm2O3} & Fisher, 99.9\%\\\hline
    Eu & Smart Elements, 99.9\% & \ch{Eu2O3} & Alfa Aesar, Reacton, 99.9\% \\\hline
    Gd & Smart Elements, 99.9\% & \ch{Gd2O3} & Rare Earth Products(?), 99.999\%\\\hline
    Tb & Smart Elements, 99.95\% & \ch{Tb4O7} & Alfa Aesar, 99.99\% \\\hline
    Dy & Alfa Aesar, 99.9\% & \ch{Dy2O3} & Alfa Aesar, 99.9\% \\\hline
    Ho & Smart Elements, 99.99\% & \ch{Ho2O3} & Alfa Aesar, Reacton, 99.99\% \\\hline
    Y & Alfa Aesar, 99.98\% & \ch{Y2O3} & Alfa Aesar, 99.999\% \\\hline
    Er & Smart Elements, 99.95\% & \ch{Er2O3} & Rare Earth Products(?), 99.99\% \\\hline
    Tm & Smart Elements, 99.9\% & \ch{Tm2O3} & * \\\hline
    Yb & Smart Elements, 99.99\% & \ch{Yb2O3} & Alfa Aesar, Reacton, 99.9\% \\\hline
    Lu & Smart Elements, 99.99\% & \ch{Lu2O3} & Alfa Aesar, Reacton, 99.995\% \\\hline
    \end{tabular}
\end{table*}

\textbf{Laboratory XRD}. Powder X-ray diffraction (pXRD) patterns were collected on a laboratory Bruker D8 Advance diffractometer with Lynxeye XE-T detector and Cu K$\alpha$ radiation ($\lambda_{K_{\alpha1}}$~=~1.5046~\AA) with Johannson monochromator in the 2$\theta$ range from 10~-~80°. Rietveld refinements on sample pXRD data with an internal silicon standard (Si powder, Acros Chem, 99.9\%) were performed using Topas 5.0 (Bruker). In all cases, the inclusion of at least 10\% bleed-through K$_{\alpha2}$ radiation ($\lambda_{K_{\alpha2}}$~=~1.5444~\AA) was found to be necessary to account for the observed minor splitting of major Bragg reflections at higher 2$\theta$.

\textbf{Raman spectroscopy}. Raman scattering spectra were collected in the backscattering geometry from 50~-~600~cm$^{-1}$ using a Horiba LabRam HR800 spectrometer with integrated 632.8 nm He-Ne laser and exchangeable  514.5 Ar ion laser (CVI Melles Griot model \#300-001, head model \#35-LAP-321-120). The spectra shown in Figure \ref{Struc_comb}d were collected with an excitation wavelength of either 632.8 nm (RE = La, Ce, Pr, Nd, Eu, Gd, Tb, Tm, Lu) or 514.5 nm (RE = Sm, Dy, Ho, Er, Yb) to minimize the fluorescence commonly observed in the optical spectra of rare-earth insulators \cite{fornasiero2004laser,cui2015raman}.

\textbf{Magnetization}. Magnetization data was collected from \textit{T}~=~1.8~-~300~K on a Quantum Design Dynacool Physical Property Measurement System (PPMS) and from \textit{T}~=~0.3~-~1.8~K on a Quantum Design Magnetic Property Measurement System (MPMS3). Magnetic susceptibility was approximated as magnetization divided by the applied magnetic field ($\chi\approx M/H$). Single-ion crystal electric field (CEF) calculations were performed from a point charge model in the \textit{J} basis using the PyCrystalField package and limited to the first coordination sphere of the RE$^{3+}$ cation \cite{scheie2021pycrystalfield}. The single-ion pseudo-spin anisotropy of each compound was extracted from calculations of the $g$-tensor components ($g_{\perp}=g_J|<-|J_-|+>|^2$, and $g_{z}=2g_J|<+|J_z|+>|^2$) where $g_J$ is the Lande g-factor, while $|+>$ and $|->$ are the CEF ground-state wavefunctions. We fixed the phase gauge of the doublet wavefunctions such that the pseudo-spin frame is aligned with the principal axes, making the $g$-tensor diagonal and strictly real. 

\textbf{Specific heat}. Specific heat data was collected on a Quantum Design Physical Property Measurement System (PPMS) using the semi-adiabatic pulse technique with a 1\% temperature rise and over three time constants. Subtraction of the estimated phonon contribution to the measured specific heat data of RE = Ce, Gd, Dy, and Yb was achieved by the scaling of the temperature scale of the measured specific heat of the relevant non-magnetic analogue by:

\vspace{-4mm}
\begin{equation*}
	\frac{\Theta^3_{\text{Mag}}}{\Theta^3_{\text{Non-mag}}} = (\frac{\text{Molar mass of}~\text{Non-mag}}{\text{Molar mass of}~\text{Mag}})^{3/2}
	\label{Ce_HC_scaling}
\end{equation*}
\vspace{-4mm}

For each non-magnetic/magnetic \ch{RE2O2S} pairing and assuming ideal stoichiometry, this value is: La/Ce~-~0.99, La/Gd~-~0.87, La/Dy~-~0.82, Lu/Gd~-~1.14, Lu/Dy~-~1.10, Lu/Yb~-~1.01. 

The change in magnetic entropy as a function of temperature, $\Delta S_{\mathrm{mag}}$, was approximated as $\Delta S_{\mathrm{mag}}$~=~$\int C_{\mathrm{P}}/T\,dT$ of the measured $C_{\mathrm{P,mag}}$ for each magnetic composition ($C_{\mathrm{P, total,mag}}$ - \text{scaled} $C_{\mathrm{P, total,non-mag}}$), from \emph{T}~=~0~-~300~K, with the entropy rise from \emph{T}~=~0~-~1.8~K being estimated from linear extrapolation over this range. 

\textbf{Synchrotron XRD}. Total scattering measurements were performed at the Brockhouse High Energy Wiggler beamline at the Canadian Light Source \cite{rahemtulla2025high}. Polycrystalline \ch{Nd2O3} samples were loaded into 0.81 mm inner diameter Kapton tubes in an Argon-filled glovebox and sealed at both ends with GE varnish. Data were collected using 65.65 keV X-rays and a Varex XRD 4343CT area detector. The pair distribution functions were produced with the GSAS-II software using a Qmax of 27 \AA$^{-1}$ \cite{toby2013gsas}.

\begin{table*}[htbp]
\centering
\caption{\textbf{Refined lattice parameters, atomic positions, and isotropic displacement parameters as a function of increasing seven-coordinate, trivalent ionic radius across the \ch{RE2O2S} series, derived from Rietveld refinement of pXRD data collected from representative \ch{RE2O2S} powder samples.} All samples were measured with an internal Si standard to ensure a reliable comparison and refined in the $P\overline{3}m1$ (164) space group.}
\begin{tabular}{|c|c|c|c|c|c|c|c|c|}
    \hline
   \multirow{3}{*}{\textbf{RE}} & \textbf{Ionic} & \multirow{2}{*}{\textbf{\textit{a}}} & \multirow{2}{*}{\textbf{\textit{c}}} & \multicolumn{2}{|c|}{\textbf{RE}} & \multicolumn{2}{|c|}{\textbf{O}} & \textbf{S}\\ 
   & \textbf{Radius} & \multirow{2}{*}{(\AA)} & \multirow{2}{*}{(\AA)} & \multicolumn{2}{|c|}{(1/3, 2/3, \textit{z})} & \multicolumn{2}{|c|}{(1/3, 2/3, \textit{z})} & (0, 0, 0) \\
   & (\AA) & & & \textbf{\textit{z}} & \textbf{$B_{eq}$} & \textbf{\textit{z}} & \textbf{$B_{eq}$} & \textbf{$B_{eq}$} \\\hline
    La & 1.1  & 4.05243(6) & 6.9466(2) & 0.2784(3) & 0.61(6) & 0.637(2) & 0.1(5) & 1.2(3) \\ \hline
    Ce & 1.07 & 4.0046(1) & 6.8757(3) & 0.2784(5) & 0.2(1) & 0.636 & 3(1) & 0.1(1) \\ \hline
    Pr & 1.058  & 3.97331(5) & 6.8266(2) & 0.2796(4) & 0.1(1) & 0.636(4) & 1.5(8) & 0.1(3) \\ \hline
    Nd & 1.046  & 3.94570(6) & 6.7878(2) & 0.2792(4) & 0.1(1) & 0.635(4) & 1.3(8) & 0.1(4) \\ \hline
    Sm & 1.02  & 3.89319(5) & 6.7152(2) & 0.2802(3) & 0.34(5) & 0.633(2) & 0.1(5) & 0.6(3) \\ \hline
    Eu & 1.01  & 3.87162(3) & 6.6858(1) & 0.2804(3) & 0.39(7) & 0.634(2) & 0.5(5) & 0.4(3) \\ \hline
    Gd & 1.00 & 3.8517(1) & 6.6648(4) & 0.2825(8) & 0.2(1) & 0.633 & 2.7(2) & 0.1(1) \\ \hline
    Tb & 0.98  & 3.82410(6) & 6.6277(2) & 0.2812(3) & 0.96(5) & 0.632(2) & 0.1(4) & 1.0(3) \\ \hline
    Dy & 0.97  & 3.80121(8) & 6.5971(2) & 0.2811(3) & 1.18(6) & 0.631(2) & 0.1(5) & 1.2(4) \\ \hline
    Y & 0.97 & 3.78412(9) & 6.5885(3) & 0.2805(4) & 2.5(1) & 0.629(2) & 0.1(4) & 2.0(3) \\ \hline
    Ho & 0.97  & 3.7811(1) & 6.5732(3) & 0.2822(6) & 1.1(1) & 0.636(4) & 1.9(9) & 0.1(1) \\ \hline
    Er & 0.945  & 3.76043(3) & 6.5485(1) & 0.2822(2) & 1.21(4) & 0.639(2) & 0.3(3) & 1.6(2) \\ \hline
    Tm & 0.937  & 3.73990(6) & 6.5210(2) & 0.2820(4) & 2.0(1) & 0.641(4) & 1.6(9) & 0.1(1) \\ \hline
    Yb & 0.925  & 3.72236(7) & 6.5002(2) & 0.2823(2) & 1.35(7) & 0.640(3) & 1.2(3) & 2.2(4) \\ \hline
    Lu & 0.919  & 3.7079(1) & 6.4893(4) & 0.2816(7) & 3.3(2) & 0.641(5) & 0.1(1) & 0.1(1) \\\hline
    \end{tabular}
    \label{pXRD_Refinement_Table1}
\end{table*}

\begin{table*}[htbp]
\centering
    \caption{\textbf{Interatomic distances and goodness of fit parameters derived from Rietveld refinement of pXRD data collected from representative \ch{RE2O2S} powder samples.} RE-\ch{O1} and RE-\ch{O2} respectively denote the distances between RE$^{3+}$ cations and O$^{2+}$ in the same plane and across the slab. \ch{RE1}-\ch{RE1} and \ch{RE1}-\ch{RE2} respectively denote nearest neighbor RE$^{3+}$~-~RE$^{3+}$ distances in the same plane and across the slab.}
\begin{tabular}{|c|c|c|c|c|c|c|c|c|c|}
    \hline
   \multirow{2}{*}{\textbf{RE}} & \textbf{\ch{RE1}-\ch{RE1}} & \textbf{\ch{RE1}-\ch{RE2}} & \textbf{RE-\ch{O1}} & \textbf{RE-\ch{O2}} & \textbf{RE-\ch{S}} & \multirow{2}{*}{\textbf{R$_\text{exp}$}} & \multirow{2}{*}{\textbf{R$_\text{wp}$}} & \multirow{2}{*}{\textbf{R$_\text{p}$}} & \multirow{2}{*}{\textbf{GOF}} \\ 
   & (\AA) & (\AA) & (\AA) & (\AA) & (\AA) & & & & \\\hline
    La & 4.0524(1) & 3.867(4) & 2.412(4) & 2.49(2) & 3.036(1) & 8.84 & 10.52 & 8.57 & 1.19 \\ \hline
    Ce & 4.0046(1) & 3.825(6) & 2.386(7) & 2.46(3) & 3.002(3)& 9.56 & 13.23 & 11.47 & 1.38\\ \hline
    Pr & 3.9733(1) & 3.784(5) & 2.365(7) & 2.43(3) & 2.984(2) & 11.64 & 12.86 & 9.70 & 1.10 \\ \hline
    Nd & 3.9457(1) & 3.765(5) & 2.351(7) & 2.42(3) & 2.963(2) & 8.61 & 9.36 & 7.84 & 1.09 \\ \hline
    Sm & 3.8932(1) & 3.710(4) & 2.322(4) & 2.37(1) & 2.931(1) & 10.45 & 12.57 & 9.02 & 1.20 \\ \hline
    Eu & 3.8716(1) & 3.690(4) & 2.307(4) & 2.36(1) & 2.917(1) & 5.80 & 6.87 & 6.18 & 1.18 \\ \hline
    Gd & 3.8517(1) & 3.654(9) & 2.294(1) & 2.34(1) & 2.914(4) & 16.2 & 19.1 & 13.9 & 1.18 \\ \hline
    Tb & 3.8241(1) & 3.645(4) & 2.282(4) & 2.33(1) & 2.889(1) & 7.08 & 8.33 & 6.61 & 1.18 \\ \hline
    Dy & 3.8012(1) & 3.627(4) & 2.270(4) & 2.31(1) & 2.873(1) & 16.91 & 18.79 & 14.18 & 1.11 \\ \hline
    Y & 3.7841(1) & 3.625(5) & 2.265(4) & 2.30(1) & 2.862(2) & 7.94 & 9.59 & 7.81 & 1.21 \\ \hline
    Ho & 3.7811(1) & 3.601(7) & 2.248(7) & 2.33(3) & 2.865(3) & 11.53 & 14.61 & 11.16 & 1.27 \\ \hline
    Er & 3.7604(1) & 3.585(3) & 2.232(4) & 2.34(1) & 2.851(1) & 6.07 & 7.72 & 5.90 & 1.27 \\ \hline
    Tm & 3.7399(1) & 3.570(5) & 2.217(6) & 2.34(3) & 2.836(2) & 10.24 & 15.69 & 13.26 & 1.53 \\ \hline
    Yb & 3.7224(1) & 3.554(4) & 2.208(5) & 2.33(2) & 2.826(1) & 4.77 & 5.83 & 4.80 & 1.22 \\ \hline
    Lu & 3.7079(1) & 3.552(8) & 2.199(8) & 2.52(4) & 2.815(3) & 11.05 & 13.26 & 10.45 & 1.20 \\\hline
    \end{tabular}
    \label{pXRD_Refinement_Table2}
\end{table*}

\begin{figure*}[ht]
    \includegraphics[width=0.5\textwidth]{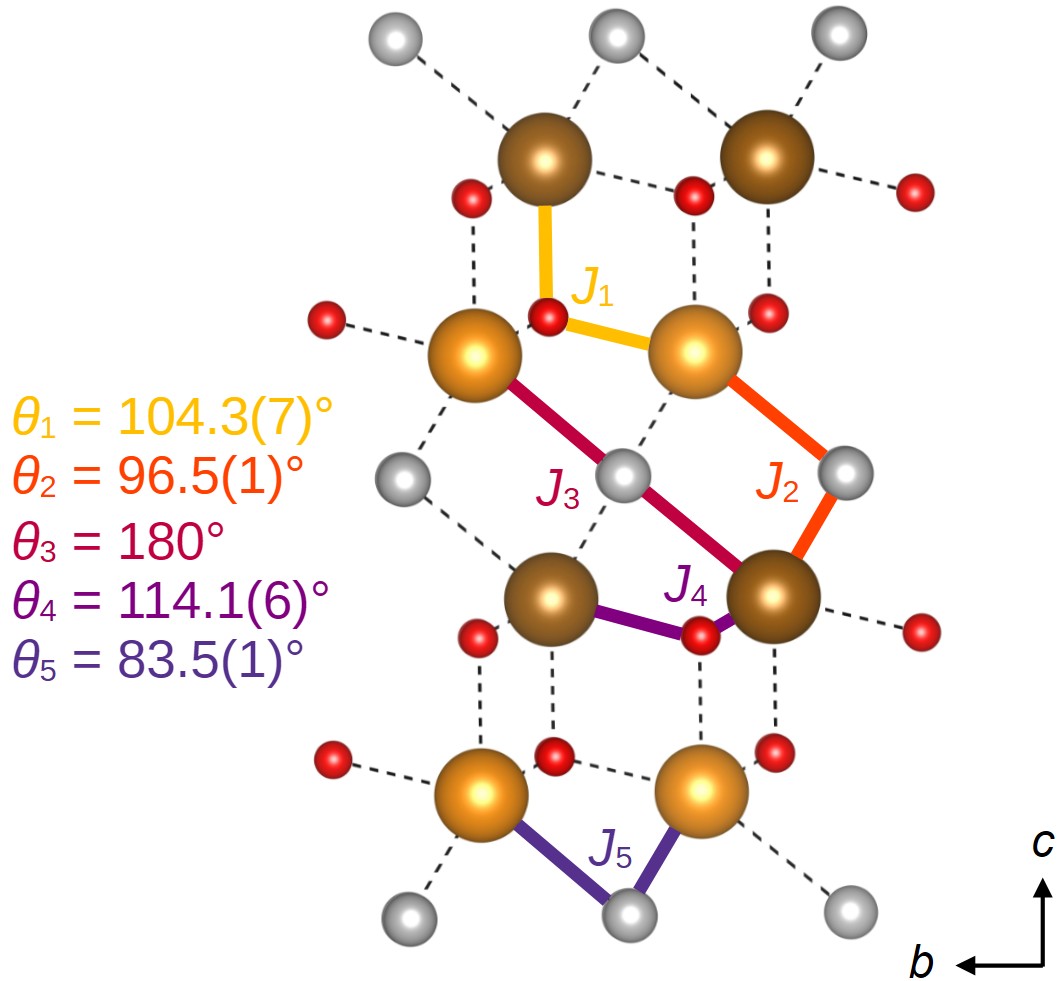}
    \caption{\textbf{Magnetic superexchange pathways in the \ch{RE2O2S} structure type.} $J_1$ and $J_4$ represent exchange through O$^{2-}$ 2\textit{p}-orbitals within the bilayer slab, while $J_2$, $J_3$, and $J_5$ represent exchange through interslab extended, hybridized S$^{2-}$ orbitals.}
    \label{Superexchange_pathways}
\end{figure*}

\begin{figure*}[ht]
    \includegraphics[width=0.5\textwidth]{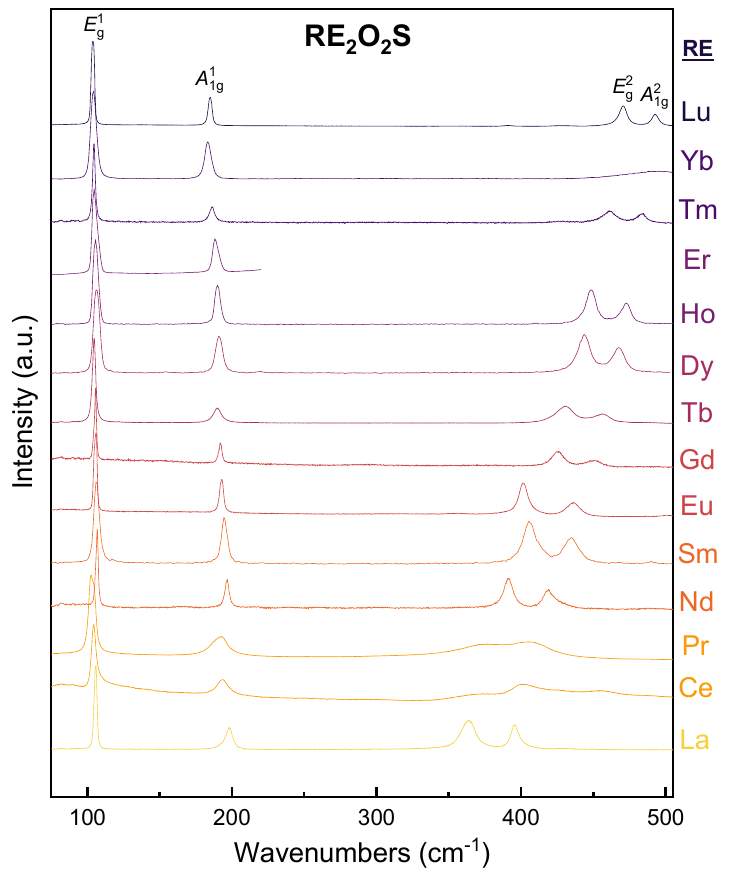}
    \caption{\textbf{Raman scattering spectra collected at room temperature from representative \ch{RE2O2S} powder samples.}}
    \label{Raman_SI}
\end{figure*}

\begin{figure*}[ht]
    \includegraphics[width=1.0\textwidth]{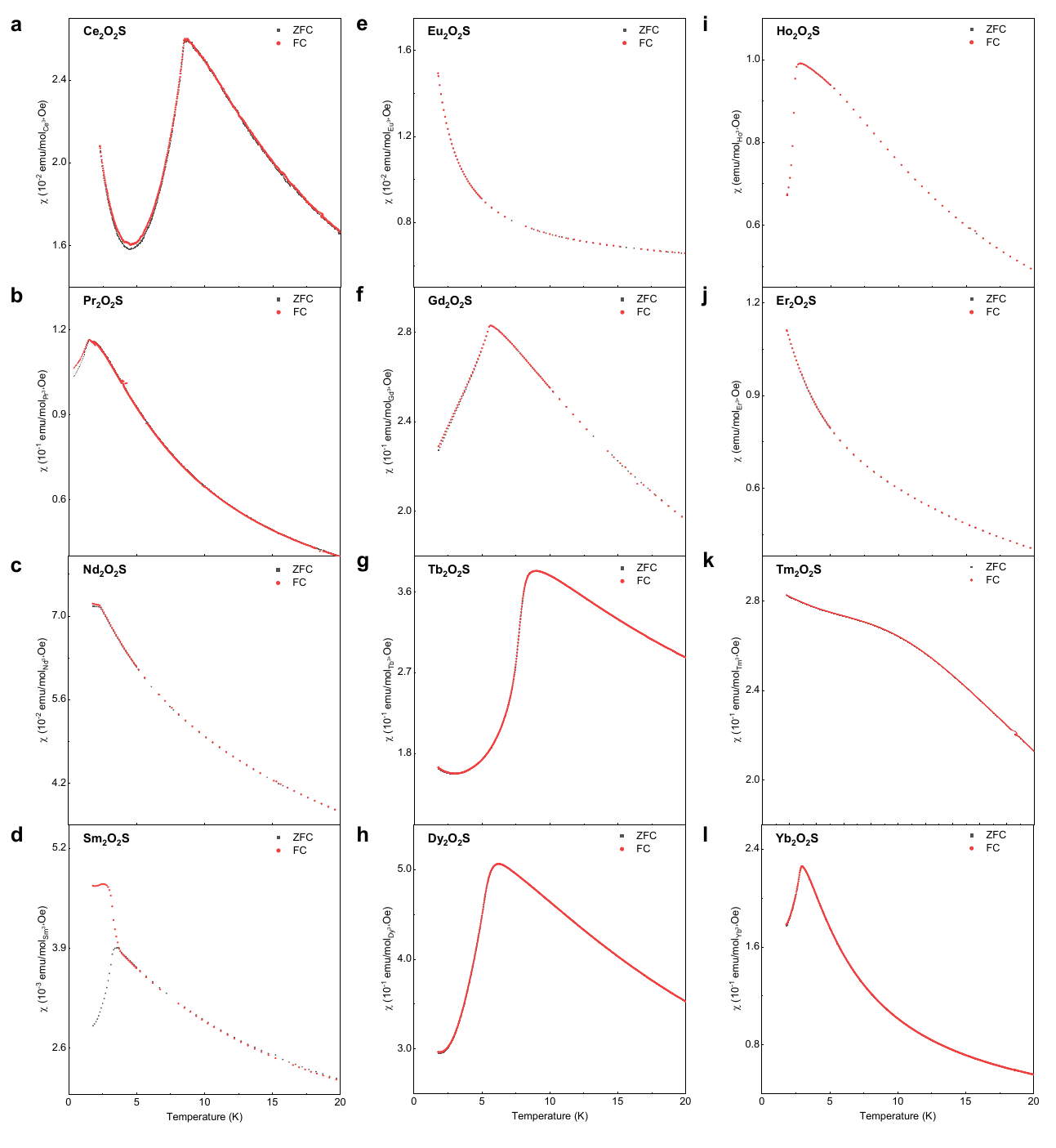}
    \caption{Magnetic susceptibility as a function of temperature measured for \ch{RE2O2S} powder samples under both zero-field-cooled (ZFC - black) and field-cooled (FC - red) conditions from \textit{T}~=~1.8~-~20~K.}
    \label{MvT_SI}
\end{figure*}

\begin{table*}
    \centering
    \caption{Fitting parameters obtained from Curie-Weiss analysis of powder magnetization measurements of representative Nd-containing phases from \emph{T}~=~5~-~10~K.}
    \label{Nd-member_lowT_CW_fits}
    \begin{tabular}{|c|c|c|c|c|c|c|}
    \hline
     \multirow{2}{*}{\textbf{Phase}} & \textbf{\emph{T}$_\text{N}$} & \textbf{$\mu_{\text{eff}}$/Nd$^{3+}$} & \textbf{$\mu_{\text{calc}}$/Nd$^{3+}$} & \textbf{$\theta_{\text{CW}}$} & \multirow{2}{*}{\textit{f}} & \multirow{2}{*}{$\chi_\text{o}$}\\ 
    & (K) & ($\mu_\text{B}$) & ($\mu_\text{B}$) & (K) & & \\\hline
    \ch{Nd2O3} & 0.55\cite{rai2020magnetism} & 3.10(1) & 3.62 & -10.2(1) & 18.5(1) & 0 \\\hline
    \ch{Nd2O2S} & 2.3 & 3.23(1) & 3.62 & -16.3(1) & 7.1(1) & 0 \\\hline
    \end{tabular}
\end{table*}

\begin{figure*}[ht]
    \includegraphics[width=1.0\textwidth]{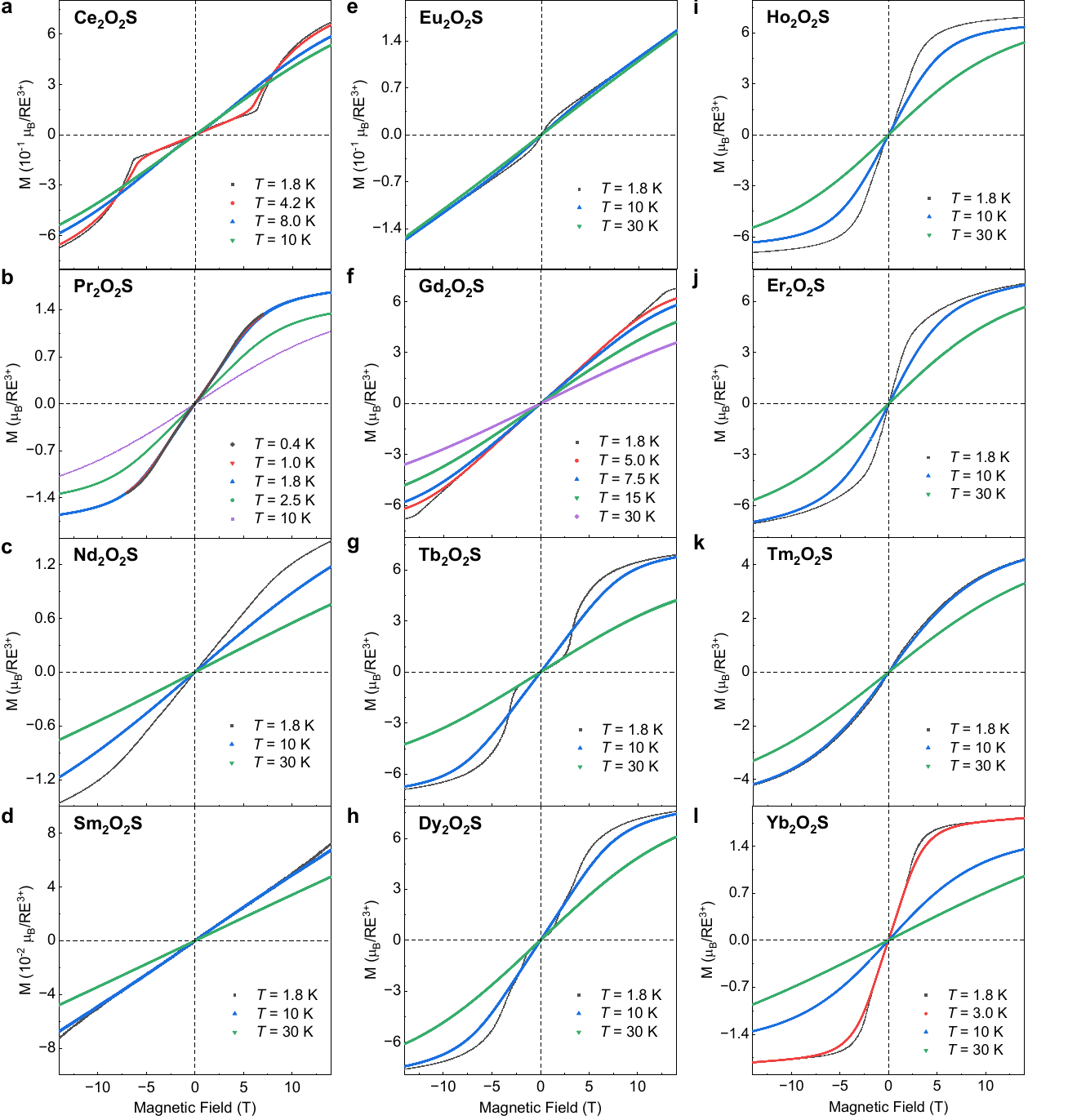}
    \caption{Field-dependent magnetization of representative \ch{RE2O2S} powder samples at various temperatures from \textit{T}~=~1.8~-~30~K.}
    \label{MvH_SI}
\end{figure*}

\begin{figure*}[ht]
    \includegraphics[width=1\textwidth]{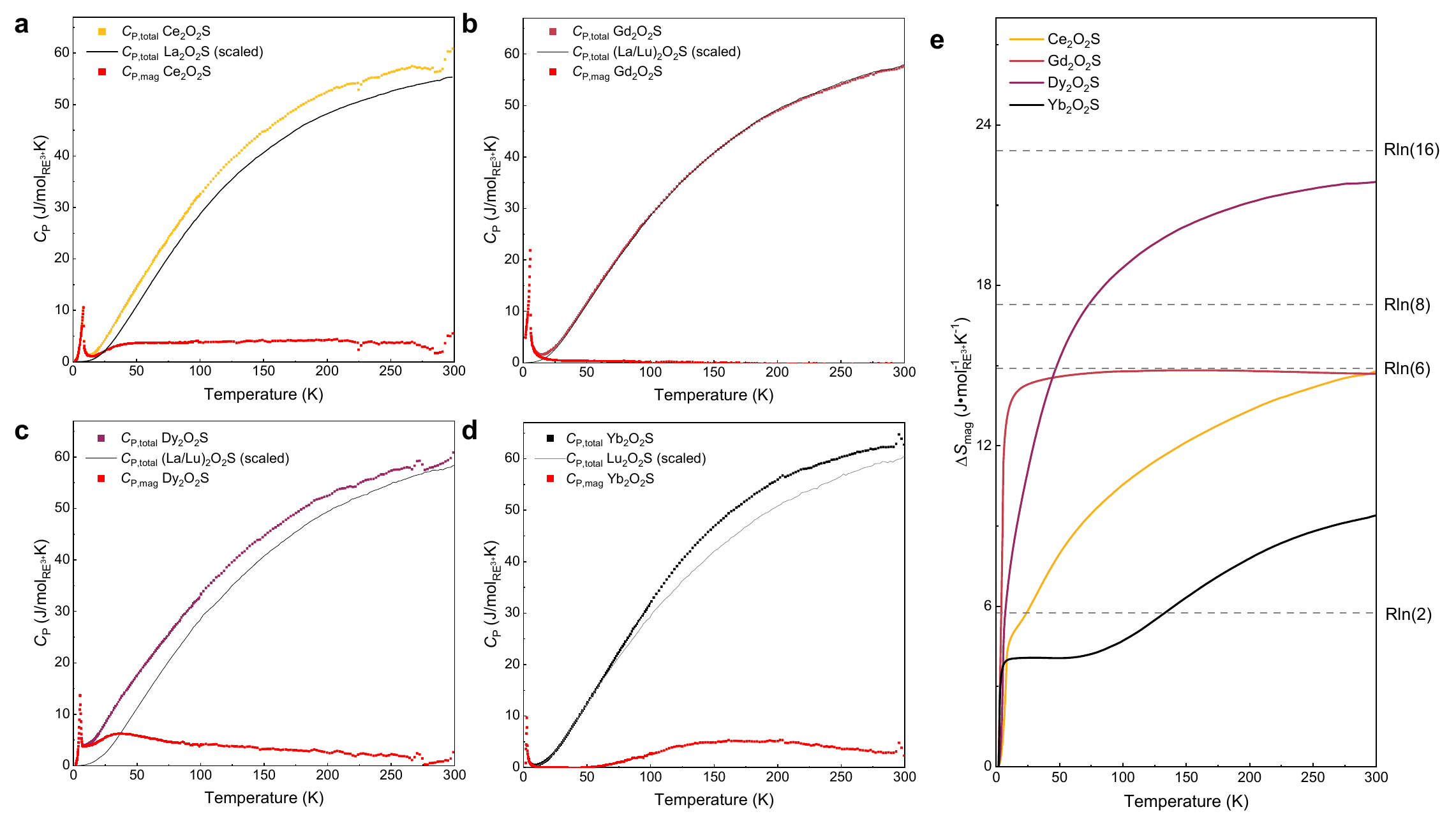}
    \caption{Specific heat as a function of temperature for (a) \ch{Ce2O2S} (gold), (b) \ch{Gd2O2S} (maroon), (c) \ch{Dy2O2S} (purple), and (d) \ch{Yb2O2S} (black) powder samples, as well as their respective non-magnetic analogues (gray). The estimated magnetic contribution to the total specific heat for each composition is shown in red. (e) Integrated magnetic entropy rise as a function of temperature for each composition up to \textit{T}~=~300~K.}
    \label{HC_SI}
\end{figure*}

\begin{figure*}[ht]
    \includegraphics[width=0.8\textwidth]{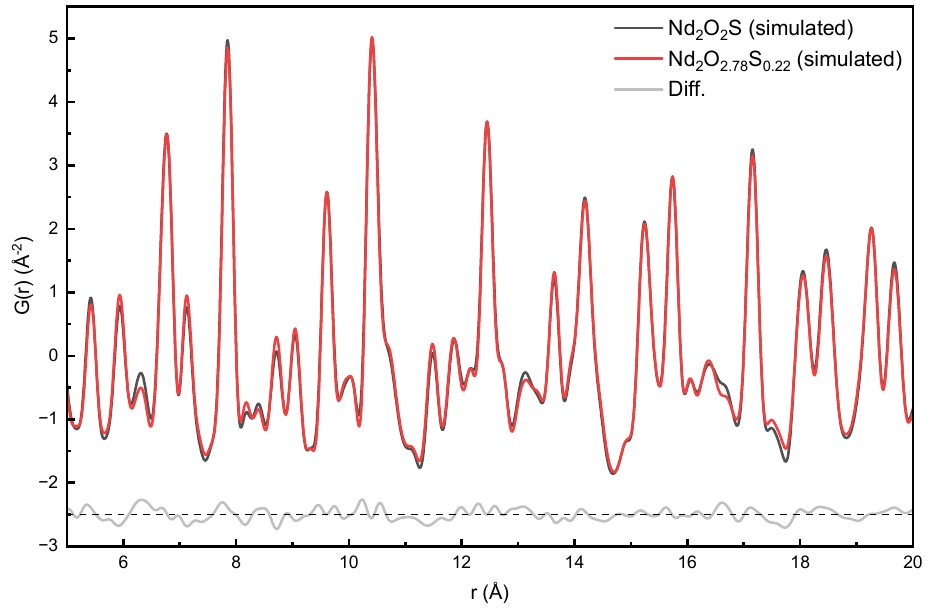}
    \caption{\textbf{Simulated reduced pair distribution function [G(r)] for stoichiometric \ch{Nd2O2S} (black) and \ch{Nd2O_{2.78}S_{0.22}} (red)} and their difference (gray), highlighting the interatomic distances most impacted by significant oxygen substitution on the sulfur site. The dotted black line represents the zero point of the difference curve (gray).}
    \label{PDF_SI}
\end{figure*}

\end{document}